\documentclass[%
 aip,
 amsmath,amssymb,
 reprint,%
]{revtex4-1}

\newcommand{\vect}[1]{\mathbf{#1}}

\usepackage{amsmath,amssymb}
\usepackage{graphicx}
\usepackage{dcolumn}
\usepackage{bm}

\usepackage[utf8]{inputenc}
\usepackage[T1]{fontenc}
\usepackage{mathptmx}
\usepackage{etoolbox}

\makeatletter
\def\@email#1#2{%
 \endgroup
 \patchcmd{\titleblock@produce}
  {\frontmatter@RRAPformat}
  {\frontmatter@RRAPformat{\produce@RRAP{*#1\href{mailto:#2}{#2}}}\frontmatter@RRAPformat}
  {}{}
}%
\makeatother
\begin{document}


\title{Sessile condensate droplets as quasi-static wall deformations\\ in direct numerical simulations of channel flow with condensation}
\author{Philipp~Bahavar}
 \email{philipp.bahavar@dlr.de}
\author{Claus~Wagner}%
\affiliation{
German Aerospace Center (DLR), Bunsenstr. 10, 37073 Göttingen, Germany%
}%
\affiliation{
Technische Universität Ilmenau, Helmholtzring 1, 98693 Ilmenau, Germany%
}%

\date{\today}

\begin{abstract}
Condensation is an important aspect of many flow applications due to the universal presence of humidity in the air at ambient conditions.
For direct numerical simulations of such flows, simulating the gas phase as a mixture characterized by temperature and humidity coupled by the release and absorption of latent heat has been shown to yield results consistent with multiphase direct numerical simulations at reduced costs.
Modeling sessile droplets as quasi-static deformations based on the underlying condensation rates at the surface extends the single-phase approach to include the interaction between droplets and flow while retaining the advantages of simulating the gas phase only.
The results of simulations of turbulent flow through a cooled, vertical channel with and without droplets show increased turbulent transport of heat and vapor due to the presence of condensate droplets, both in the immediate vicinity of the droplets and for the channel as a whole.
In particular, the breaking of the symmetry of the wall shifts condensation from the surrounding areas onto the droplet surfaces.

\end{abstract}

\maketitle

\section{Introduction}
Flows of fluids with condensable components are relevant to a wide array of applications, from the design of heat exchangers to the formation of clouds in the atmosphere.

The universal presence of humidity in the ambient air means that condensation can occur in any space where humid air is exposed to temperature differences.
This condensation can present a problem, such as for example the fogging of the windshield in a car, where the warm air in the heated interior meets the cool glass surface.
The energy required for defogging the windscreen is considerable, especially considering electric vehicles that cannot easily utilize waste heat of the engine.\cite{Lorenz2015}

The combination of thermodynamic and fluid-mechanical effects and phenomena presents a challenge when trying to model flows with phase transitions due to the variety of interactions and the broad range of length and time scales involved.
For application-oriented research, models based on the relative humidity and average mass transfer rates are applied to Reynolds-averaged Navier-Stokes (RANS) simulations, which provide large-scale results for condensation and evaporation on surfaces.\cite{Leriche2015}

For investigations of the fundamental mechanisms and interactions between flow and phase transitions, fully resolved direct numerical simulations (DNS) utilize a spectrum of techniques to calculate multiphase flows and phase transition in such flows.
They range from approaches with inert, ideal tracer particles, one- or two-way coupling between fluid and particles via drag, four-way coupling additionally including particle--particle collisions, to methods with fully resolved interfaces and internal dynamics.
For an overview, refer for example to reviews by Balachandar and Eaton\cite{Balachandar2010} or Kuerten.\cite{Kuerten2016}
In general, the high costs of multiphase DNS tend to limit the investigations to generic geometries\cite{Russo2014} or individual droplets.\cite{Orazzo2019}

For flows at ambient conditions, where vapor loads and temperature differences are limited compared to technical applications, condensation is almost exclusively confined to the cooled surfaces themselves, with only negligible condensation within the volume.
In this case, using a single-phase DNS of a homogeneous mixture of dry air and water vapor and modeling phase transition effects with sources and sinks for the vapor concentration and associated latent heat yields reliable results as validated against two-way coupled multiphase DNS.\cite{Bahavar2020}
The concentration of condensation on surfaces allows an extension of the approach to include sessile droplets in the single-phase simulation.
The interaction between droplets and flow is complementary to the one-way coupling of fluid and particles, where the particle movement is governed by the fluid flow.
Instead, static droplets are modeled at the surface, where they interact with and influence the flow, while, once placed, the droplets are static in their position and shape, independent of the instantaneous flow situation.

On top of the effect of the condensate droplets, thermodynamic effects of condensation can still be included in the single-phase formulation via the coupling between latent heat of condensation, temperature, vapor concentration, and thermal and solutal buoyancy.

The accuracy of the single-phase DNS paired with the comparatively lower computational cost for including phase transitions make this approach well-suited to investigate the interaction between condensation and turbulent flow, which necessitates a comparatively large flow domain.\cite{Jimenez1991}

This article presents the application of the extended approach to the turbulent flow of humid air through a cooled, vertical channel geometry, with flow parameters chosen to represent conditions relevant to vehicle or cabin ventilation.

\section{Governing equations}
The approach to treating condensation in the context of vehicle ventilation leverages the specific thermodynamic parameters specific to this application.
Humid air is described as a homogenous fluid, comprising a carrier phase of dry air, into which water vapor is dissolved.\cite{Bahavar2020}
The dynamics of this fluid with density $\rho$ and kinematic viscosity $\nu$ are described by the Navier-Stokes equations for incompressible flow of a fluid with constant properties,
\begin{align}
  \nabla \cdot \vect{u} = 0,\label{eq:continuity}\\
  \frac{\partial \vect{u}}{\partial t} + (\vect{u}\cdot \nabla) \vect{u} = -\frac{1}{\rho}\nabla p + \nu \nabla^2\vect{u} + \vect{f},\label{eq:velocity}
\end{align}
where eq. (\ref{eq:continuity}) provides the continuity condition and eq. (\ref{eq:velocity}) the evolution of the velocity $\vect{u}$ experiencing a pressure gradient $\nabla p$ and a general body force $\vect{f}$.

Both temperature $T$ and humidity in the form of molar vapor concentration $c$ of the humid air are described by convection-diffusion equations with general source terms $S$,
\begin{align}
  \frac{\partial T}{\partial t} + (\vect{u}\cdot \nabla)T =& \kappa \nabla^2 T + S_T,\label{eq:temperature_cd}\\
  \frac{\partial c}{\partial t} + (\vect{u}\cdot \nabla)c =& D \nabla^2 c + S_c,\label{eq:concentration_cd}
\end{align}
with thermal diffusivity $\kappa$ and the binary mass diffusion coefficient $D$.

The presence of temperature and humidity gives rise to thermal and solutal buoyancy $\vect{B}_T$ and $\vect{B}_c$, respectively.\cite{Hammou2004}
For the ranges considered here, the Boussinesq approximation applies\cite{Gray1976} and the linearized buoyant forces are
\begin{align}
  \vect{B}_T =& -\beta_T (T - T_{\mathit{ref}})\vect{g}\label{eq:Boussinesq_T}\\
  \vect{B}_c =& -\beta_c (c - c_{\mathit{ref}})\vect{g}\label{eq:Boussinesq_c}
\end{align}
with the gravitational acceleration $\vect{g}$.
The expansion coefficients with respect to temperature and concentration are
\begin{align}
  \beta_T = \left.\frac{1}{\rho}\frac{\partial \rho}{\partial T}\right\rvert_{T_\mathit{ref}},\quad \beta_c = \left.\frac{1}{\rho}\frac{\partial \rho}{\partial c}\right\rvert_{c_\mathit{ref}}.\label{eq:expansion_coeffs}
\end{align}
Setting the force in eq. \ref{eq:velocity} to
\begin{align}
  \vect{f} = \vect{B}_T + \vect{B}_c
\end{align}
includes the buoyancy contributions in the velocity equation.
Thus, temperature and vapor concentration act as active scalars in this framework, coupled to the velocity field via buoyancy and in turn affected by the convective transport with the flow.

Considering first the effect of condensation on the gas phase, the source term $S_c$ in eq. (\ref{eq:concentration_cd}) expresses the loss or gain of vapor concentration in a volume $V$ due to condensation or evaporation across a surface $A$, using the ideal gas law as the equation of state and the Hertz--Knudsen--Schrage equation\cite{Marek2001}
\begin{align}
  S_c = -\frac{2\sigma_c}{2-\sigma_c}\frac{A}{V}\sqrt{\frac{RT}{2\pi M}}(c-c_\mathit{sat}).\label{eq:HKS_source}
\end{align}
Here, $\sigma_c$ is the condensation coefficient, expressing the probability that a vapor molecule stays in the liquid phase after condensation, and is assumed to be equal to the evaporation coefficient in this formulation.
The condensation rate is proportional to the difference between the vapor concentration to the saturation value $c_\mathit{sat}$.
This saturation concentration is itself a function of the temperature, described for example by the Magnus formula.\cite{Alduchov1996}

Conversely, the temperature source term $S_T$ in eq. (\ref{eq:temperature_cd})
captures the effect of heating or cooling from the release or absorption of latent heat during the phase transition,
\begin{align}
  S_T = -\frac{h_v}{c_p} S_c,\label{eq:temperature_source}
\end{align}
with the latent heat of vaporization $h_v$ and the specific heat capacity of the fluid $c_p$.
This connection between temperature and concentration completes the full coupling between the two scalar fields and the velocity.

The treatment of liquid water resulting from condensation is separate for liquid inside the volume and on cooled surfaces.
With the low overall vapor concentrations and limited subcooling encountered in vehicle ventilation, the amount of condensate forming inside the volume is negligible.
Consequently, it is discarded in this approach, and no transport of liquid water with the flow of humid air is simulated.

Instead, condensation occurs predominantly on cooled surfaces.
In wall-bounded flows, droplets forming on these walls have the potential to significantly affect the flow throughout the whole geometry.
This surface modification presents an additional mechanism for condensation to influence such flows.

Condensate mass accumulates at the surface at the same rate as it is removed from the fluid by condensation, expressed by the source term in eq. (\ref{eq:HKS_source}).
In the upper limit of  complete condensation of all available vapor, the mass flux across the phase boundary $\phi_\mathit{PT}$ is equal to the vapor mass flux convected by the fluid
\begin{align}
  \phi_c = c \rho_v u_b,\label{eq:convective_flux}
\end{align}
where $u_b$ is the convection velocity and $\rho_v$ the density of pure vapor.
Assuming that the condensate grows as a uniform film across a cooled surface, the film thickness grows with velocity
\begin{align}
  u_f = \phi_\mathit{PT} \, \frac{1}{\rho_l} \leq \frac{c \rho_v}{\rho_l}\, u_b.\label{eq:film_growth_velocity}
\end{align}
Since the ratio between the vapor density $\rho_v$ and the liquid density $\rho_l$ is very small, and the vapor concentration $c$ at ambient conditions is small as well, the theoretical film growth velocity $u_f$ is negligible compared to the convection $u_b$.\cite{Tryggvason2005}
Consequently, momentum transfer at the fluid-liquid interface is negligible, and the interface is effectively static with respect to the flow.

The slow condensate growth makes fully resolved simulations over the time scales necessary to reach appreciably droplet sizes unfeasible.
Instead, the long time scales of droplet growth allow for extrapolation based on the time-averaged local condensation rates.
To this end, the accumulated condensate mass is estimated as
\begin{align}
  m_c = \left\langle \phi_\mathit{PT} \right\rangle\,t_g,\label{eq:mass_estimate}
\end{align}
where $t_g$ is the desired growth time.

\begin{figure}[h]
\includegraphics[width=\columnwidth]{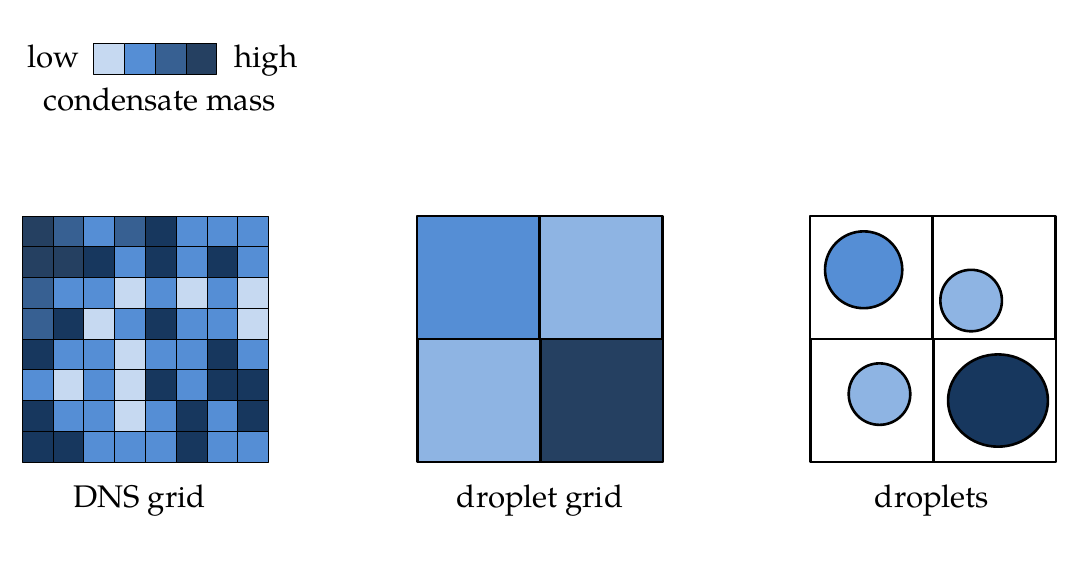}
\caption{Illustration of the grid-level mass conservation of the droplet pattern. Extrapolated condensate masses from the surface cells of the DNS are accumulated in cells with sizes on the order of the droplet radii, and then combined into droplets with volumes corresponding to the available mass. \label{fig:droplet_generation}}
\end{figure}

The microscopic mechanisms of nucleation, which necessitate simulations on molecular level,\cite{Niu2018} and the concurrent isolated droplet growth and droplet coalescence,\cite{Medici2014} which span length and time scales orders of magnitude beyond those of the turbulent flow, preclude a fully resolved simulation of the complete dynamics.
To avoid these complications, the available condensate mass after $t_g$ is consolidated into synthetic droplets, whose effect on the turbulent flow is then investigated.
A grid is applied to the surface, with cell sizes on the order of the average footprint radius of droplets with the extrapolated mass.
Hemispherical droplets are then generated according to the condensate mass of the specific grid cell and placed randomly within the cell to avoid artificial regularity.
This approach preserves the inhomogeneity of the average condensation rate across the cooled surface, thus conserving mass on the grid scale.
An illustration of this procedure is given in Figure \ref{fig:droplet_generation}.

Since the existence of droplets changes the flow across the surface, the average condensation rate is affected as well.
Accordingly, the growth time $t_g$ needs to be short enough that the majority of the skipped growth dynamics are still governed by the initial values for the average condensation.

Owing to the effectively static nature of the droplets, this investigation focusses on their influence on the flow via the modification of the cooled surface.
For a Reynolds number of $\mathit{Re}=2000$ and subcooling on the order of $10\,\mathrm{K}$, the minimal size for a thermodynamically viable droplets is five orders of magnitude smaller than the Kolmogorov scale.
Consequently, the large majority of the droplets are far smaller than the smallest scales of the turbulent flow.\cite{Kim2011}
Restricting the investigation to only the largest droplets circumvents the necessity to resolve droplets across all scales of the self-similar growth process.
The droplets are modeled as solid, hemispherical wall deformations, disregarding their internal properties, such as the effect of curved liquid interfaces on the saturation pressure or the existence of a temperature gradient within a droplet.\cite{Rose2002}
This approach isolates the direct geometric effect of the droplets on the flow and subsequently on the transport of heat and vapor, without confounding thermodynamic effects.

\section{Simulation setup}
The governing equations are solved using the finite-volume code OpenFOAM.\cite{OpenFOAM}
The time integration uses an explicit, second-order accurate Euler--leapfrog scheme,
\begin{align}
  \vect{u}_{t+1} = \vect{u}_{t-1} + 2\Delta t \left(D_{t-1} + C_t + P_t\right),
\end{align}
with a time-lagged diffusion term $D_{t-1}$,\cite{Manhart2004} convection term $C_t$, global pressure gradient $P_t$, and time step $\Delta t$.
An averaging step is introduced every 50 iterations to avoid unphysical oscillations.
The projection method\cite{Chorin1968} couples pressure and velocity for the incompressible flow by solving the Poisson equation for the pressure and correcting the initial guess for the updated velocity, ensuring a divergence-free field.
Second-order central differences are used for the spatial discretization.\cite{Kath2016}

The investigated geometry is a channel with half-width $\delta$ and overall dimensions $6\pi\delta \times 2\delta \times 2\pi\delta$.
It is oriented vertically, with flow from top to bottom along the direction of gravity.
At both ends, the boundaries create an inlet-outlet configuration, while periodic boundaries apply in the spanwise direction.

Humid air at ambient conditions is chosen as the working fluid, with Prandtl number $\mathit{Pr}=\nu/\kappa=0.73$ and Schmidt number $\mathit{Sc}=\nu/D=0.65$.

The temperature $T_\mathit{in}$ and the vapor concentration $c_\mathit{in}$ at the inlet are chosen such that the air is undersaturated, $c_\mathit{in} < c_\mathit{sat}(T_\mathit{in})$.
Expressed with the dew point temperature, this is equivalent to $T_\mathit{dp}(c_\mathit{in}) < T_\mathit{in}$.
One of the channel walls is adiabatic, while the opposing wall is cooled to $T_c = T_\mathit{in}-\Delta T$.
This wall temperature is such that the subcooling with respect to the inlet vapor concentration is $T_\mathit{dp}(c_\mathit{in}) - T_c = 0.38\Delta T$, allowing condensation to occur.
The concentration at the cooled wall evolves freely and is entirely determined by the temperature, resulting in a lower bound of $c_\mathit{sat}(T_c)$.

The thermal Grashof number quantifies the buoyancy due to the temperature difference $\Delta T = T_\mathit{in} - T_c$,
\begin{align}
  \mathit{Gr}_T = \frac{g \beta_T \Delta T \delta^3}{\nu^2} = 38\,000.\label{eq:Grashof_T}
\end{align}
Analogously, the solutal Grashof number is determined by $\Delta c = c_\mathit{in} - c_\mathit{sat}(T_c)$,
\begin{align}
  \mathit{Gr}_c = \frac{g \beta_c \Delta c \delta^3}{\nu^2} = 1500.\label{eq:Grashof_c}
\end{align}
Both buoyancy contributions act in the same direction.
A decrease in temperature leads to a downwards force, same as a decrease in vapor concentration.
Together with a bulk velocity $u_b$, resulting in bulk and friction Reynolds numbers
\begin{align}
  \mathit{Re} = \frac{u_b \delta}{\nu} = 2000,\quad \mathit{Re}_\tau = \frac{u_\tau \delta}{\nu} = 135,\label{eq:Reynolds_numbers}
\end{align}
the balance between natural and forced convection is expressed via the Richardson number,
\begin{align}
  \mathit{Ri} = \frac{\mathit{Gr}_T + \mathit{Gr}_c}{\mathit{Re}^2} = 0.01.\label{eq:Richardson_number}
\end{align}
Finally, the Jakob number $\mathit{Ja}=\Delta T c_p/h_v=0.012$ gives the ratio
between sensible and latent heat in the coupling between temperature and concentration via the phase transition.

\begin{figure}[h]
\includegraphics[width=\columnwidth]{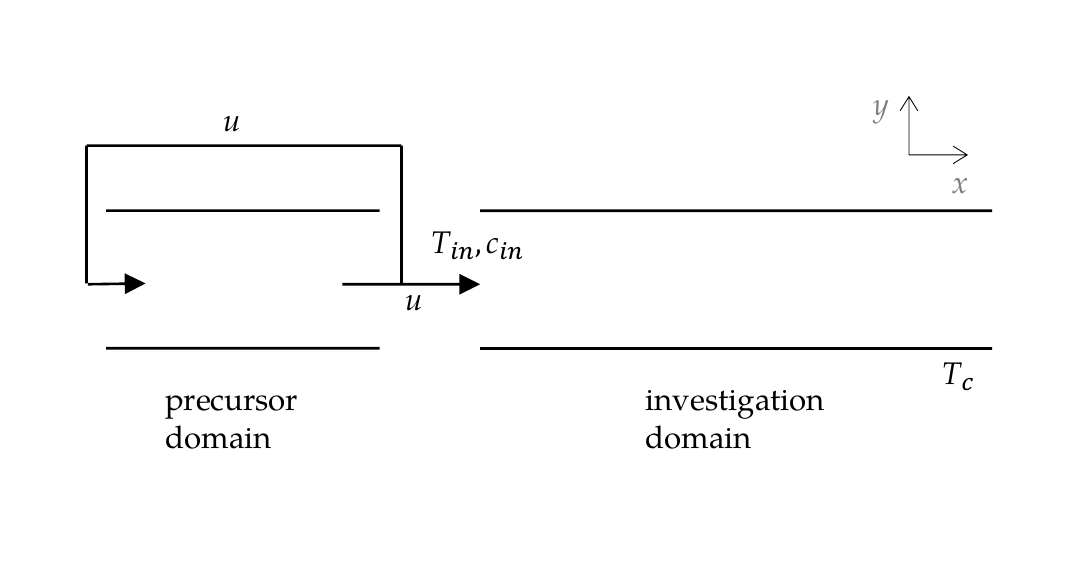}
\caption{A schematic overview of the coupled simulation setup, consisting of a cyclic precursor domain generating isothermal, turbulent flow, and the investigation domain in an inlet-outlet configuration with a cooled wall. \label{fig:simulation_schematic}}
\end{figure}

To investigate fully developed turbulent flow consistent with the desired Reynolds number along the complete length of the channel, the velocity inlet conditions are provided by an auxiliary domain.\cite{Lund1998,Bellec2017}
It consists of a biperiodic channel with dimensions $4\pi\delta \times 2\delta \times 2\pi\delta$, matching wall distance and width to the primary channel.
Inside this channel, isothermal turbulent flow with bulk velocity $u_b$ is continuously simulated.
At the cyclic outlet, the velocity field is mapped back to the inlet, but also fed forward to the inlet boundary of the investigation domain.
In this way, the generation domain serves as an infinite-length inflow channel, allowing all turbulence statistics to develop fully before entering the investigation domain.
Figure \ref{fig:simulation_schematic} gives a schematic overview of this setup.

The pressure boundaries complement the velocity boundaries, with a zero-gradient condition at the inlet and fixed pressure at the outlet, matching the zero-gradient outlet condition applied to the velocity.

To satisfy the resolution requirements of the DNS, the channel is discretized into $480 \times 180 \times 316$ hexahedral cells.
The distribution is uniform along the streamwise and spanwise directions, resulting in a resolution of $\Delta x^+ = 5.3$ and $\Delta z^+ = 2.7$, respectively.
In the wall-normal direction, the resolution is increased, with 15 grid points within the viscous sublayer at the channel walls to adequately capture the high gradients present here.
Using a hyperbolic tangent distribution, the resolution varies between $\Delta y^+ = 0.2$\,--\,$3.2$, with maximum resolution directly at the walls and minimum resolution in the center of the channel.

In the generation domain, the mesh resolution is identical, resulting in $320 \times 180 \times 316$ cells.
Exact matching of the meshes of both domains avoids numerical artifacts at the boundary.

The simulation is initialized by calculating isothermal flow with the prescribed Reynolds number until statistically steady turbulence is established in both domains.
Subsequently, the calculations for the evolution of the temperature (\ref{eq:temperature_cd}) and vapor concentration (\ref{eq:concentration_cd}) are included.
The simulation then continues until a statistically steady state accounting for influence of the added scalar fields is reached.
In this configuration, data is then collected over an interval $T=30\,\delta/u_\tau = 30\,t^+$ both to obtain the average condensation rates needed for the creation of the droplet pattern as well as to serve as a basis for comparison for the following simulation including these droplets.

\begin{figure}[h]
\centering
\includegraphics[scale=1]{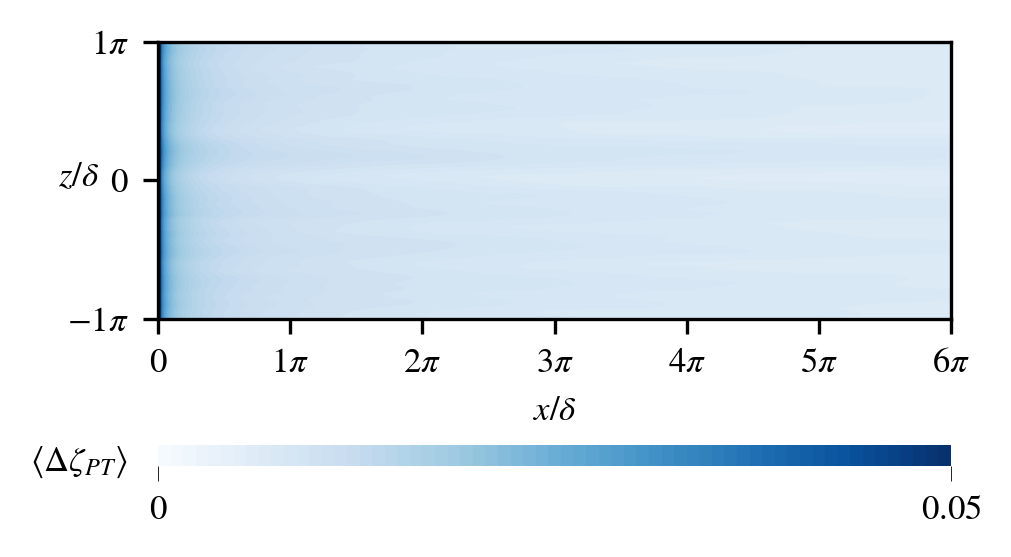}
\caption{The distribution of the average condensation rate at the cooled wall of the smooth channel. Very high rates are found directly at the inlet. Condensation diminishes along the channel, and spanwise variation can be observed.\label{fig:mdot_smooth}}
\end{figure}

Figure \ref{fig:mdot_smooth} shows this average condensation rate directly at the cooled wall expressed via the dimensionless concentration,
\begin{align}
  \zeta = \frac{c - c_\mathit{sat}(T_c)}{\Delta c},\label{eq:zeta_definition}
\end{align}
The condensation intensity is inhomogeneous across the surface.
Condensation rates are highest directly at the inlet of the channel, where fluid carrying the full vapor load first encounters the cooled wall.
As the streamwise position increases, the progressive drying of the fluid from condensation upstream reduces the driving force for further condensation, and the rates drop accordingly.
This streamwise evolution is combined with periodic spanwise variation showing the impact of the structure of the turbulent flow on the convective transport of vapor to the wall.

Extrapolating the average condensation rate necessitates a target droplet size to fix the condensate mass $m_c$ and thereby the assumed growth time $t_g$ in eq. (\ref{eq:mass_estimate}).

Since the focus of this investigation lies in the interaction between the turbulent flow and the geometric deformation of the wall, while first- and second-order statistical moments are largely unaffected by small roughness elements within the viscous sublayer,\cite{Chan-Braun2011} $t_g$ is chosen such that the smallest droplets reach a height of $h^+=5$.
In this way, all droplets have the potential to influence the flow through the channel, thus providing maximal data on the investigated interactions.

\begin{figure}[h]
\centering
\includegraphics[scale=1]{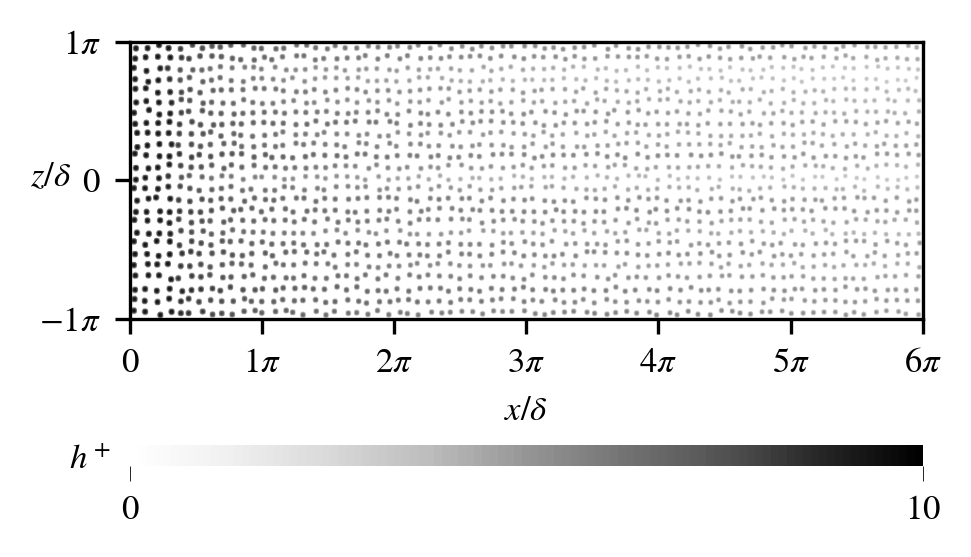}
\caption{The droplet pattern placed on the cooled wall, with different droplet heights and footprints ensuring local conservation of condensate mass.\label{fig:droplet_pattern}}
\end{figure}

The resulting pattern consists of $N=1900$ spherical caps, with a mean height of $\langle h^+\rangle_N = 6.4$ and a mean footprint radius $\langle R\rangle_N=8.9$.
Figure \ref{fig:droplet_pattern} gives an overview of the pattern and the distribution of the droplet sizes across the cooled wall.
Consistent with the underlying condensation rates, droplets tend to be larger near the inlet, with sizes decreasing along the streamwise direction.
Droplet volumes are adjusted by reducing the droplet height and footprint while keeping the radius of curvature constant.
For the largest droplets, this yields perfect hemispheres with $h_\mathit{max}^+=R_\mathit{max}^+=9.5$ and a contact angle of $\vartheta=90^\circ$, while the mean contact angle is $\langle\vartheta\rangle_N=71^\circ$.
These droplet shapes represent parts of the growth dynamics, where coalescence of droplets reduces the contact angle of the resulting droplet, which then gradually increases again with subsequent growth from additional condensation.

With the surface tension $\sigma$, both the Bond number
\begin{align}
  \mathit{Bo}=\frac{\rho_l g R_\mathit{max}}{\sigma}=0.42\label{eq:Bond_number}
\end{align}
and the Weber number
\begin{align}
  \mathit{We}=\frac{\rho u_b^2 R_\mathit{max}}{\sigma}=0.007\label{eq:Weber_number}
\end{align}
are small, and deviations from the spherical shape due to the influence of gravity and the drag of the impinging flow are negligible.

The droplets are included in the simulation as static deformations of the cooled channel wall.
To ensure that the droplet surfaces are resolved by at least six grid points along the streamwise and spanwise directions to adequately capture their curvature, the mesh in the investigation region is refined in the streamwise direction, with $\Delta x^+ = \Delta z^+=3$.
At the boundary of the generation domain, the grid spacing is matched to the investigation domain, but a hyperbolic tangent distribution is applied in the streamwise direction to reduce the number of cells in the precursor by $36\%$ compared to a uniform, overresolved mesh.
Consequently, the resolution varies between $\Delta x^+=3$ at the cyclic boundaries and $\Delta x^+=6$ in the center.
In total, the refined mesh consists of $(360\,+\,840)\times 162 \times 280$ cells across both domains.

To preserve the integrity and quality of the mesh, the deformation at the wall surface is transferred into the volume.
The shift of each volume grid point based on the surface deformation is weighed using the Wendland function $\mathrm{W2}$,
\begin{align}
  \mathrm{W2}(r) =%
  \begin{cases}%
    \left\lvert r-r_\mathit{rbf}\right\rvert \left(1 - \frac{r}{r_\mathit{rbf}})\right)^4 \left(1 + 4\frac{r}{r_\mathit{rbf}}\right),\quad r \leq r_\mathit{rbf},\\
    0\quad \mathrm{else},%
    \end{cases}
\end{align}
with base radius $r_{rbf}^+=65$ as a radial base function.
Solving a density matrix equation for the influence of multiple surface deformation per grid point inside the volume ensures a smooth adaptation of the complete mesh to the new surface.\cite{Koethe2014}

As before, the simulation is run with the new boundary conditions until a statistically steady state is reached, after which statistical data on the flow fields and their fluctuations are collected until fully converged.

\section{Local modifications of the flow fields around droplets}
\begin{figure}[h]
\centering
\includegraphics[scale=1]{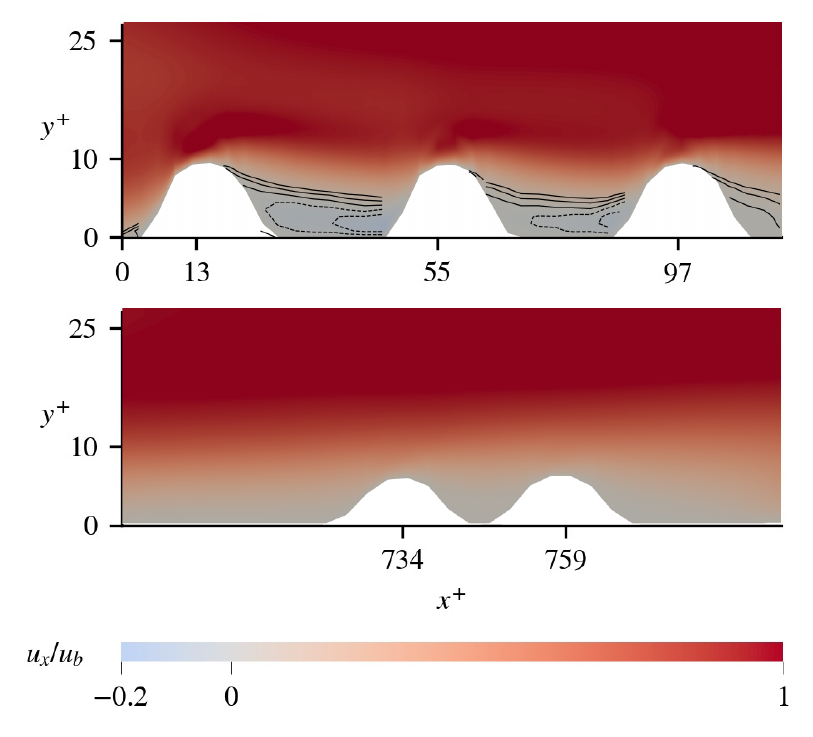}
\caption{Top panel: the instantaneous streamwise velocity in a $z$-normal plane through droplets near the channel inlet shows separation and recirculation behind the droplets. Contours in the relevant regions are spaced by $0.05$ for $u_x$ between $-0.1$ and $0.1$.\\
Bottom panel: further downstream, the modifications of the flow due to droplets upstream reduce the immediate influence of the droplets on the instantaneous velocity. \label{fig:droplet_snaps}}
\end{figure}

Figure \ref{fig:droplet_snaps} shows the direct interaction between the droplets and the instantaneous flow field at two selected locations.
The top panel shows the situation near the channel inlet, where the flow from the smooth precursor domain first encounters the cooled and deformed wall.
The droplets strongly modify the previously undisturbed flow, with separation visible at the crests and recirculation of the fluid in the wake region.

Further along the channel, the changes to the flow from the large number of droplets upstream reduce the immediate effect of additional droplets downstream.
The fluid flows around the droplets, but no separation or recirculation occurs.

Leveraging the large number of droplets present in the simulation, averaging equivalent points on or in the vicinity of the deformation over the ensemble of all droplets allows the investigation of the effect of an average droplet on the flow.
In the neighborhood of a droplets $i$, its footprint radius $R_i$ and the droplet height $h_i$ are the characteristic length scales in the $x-z$ plane and in the wall-normal $y$ direction, respectively.
To identify equivalent points between different droplets, droplet-specific cylindrical coordinates are defined, using the radial distance $r$ in the $x-z$ plane from the droplet center normalized with $R_i$, the height $y$ over the undisturbed surface normalized with $h_i$, and the angle $\varphi$ with respect to the mean flow direction along the channel axis.
Points with the same coordinates are then comparable across all droplets.
The average of a variable over the ensemble of $N$ droplets on top of the temporal average is denoted by $\langle\cdot\rangle_N$ in the following.

\begin{figure}[h]
\centering
\includegraphics[scale=1]{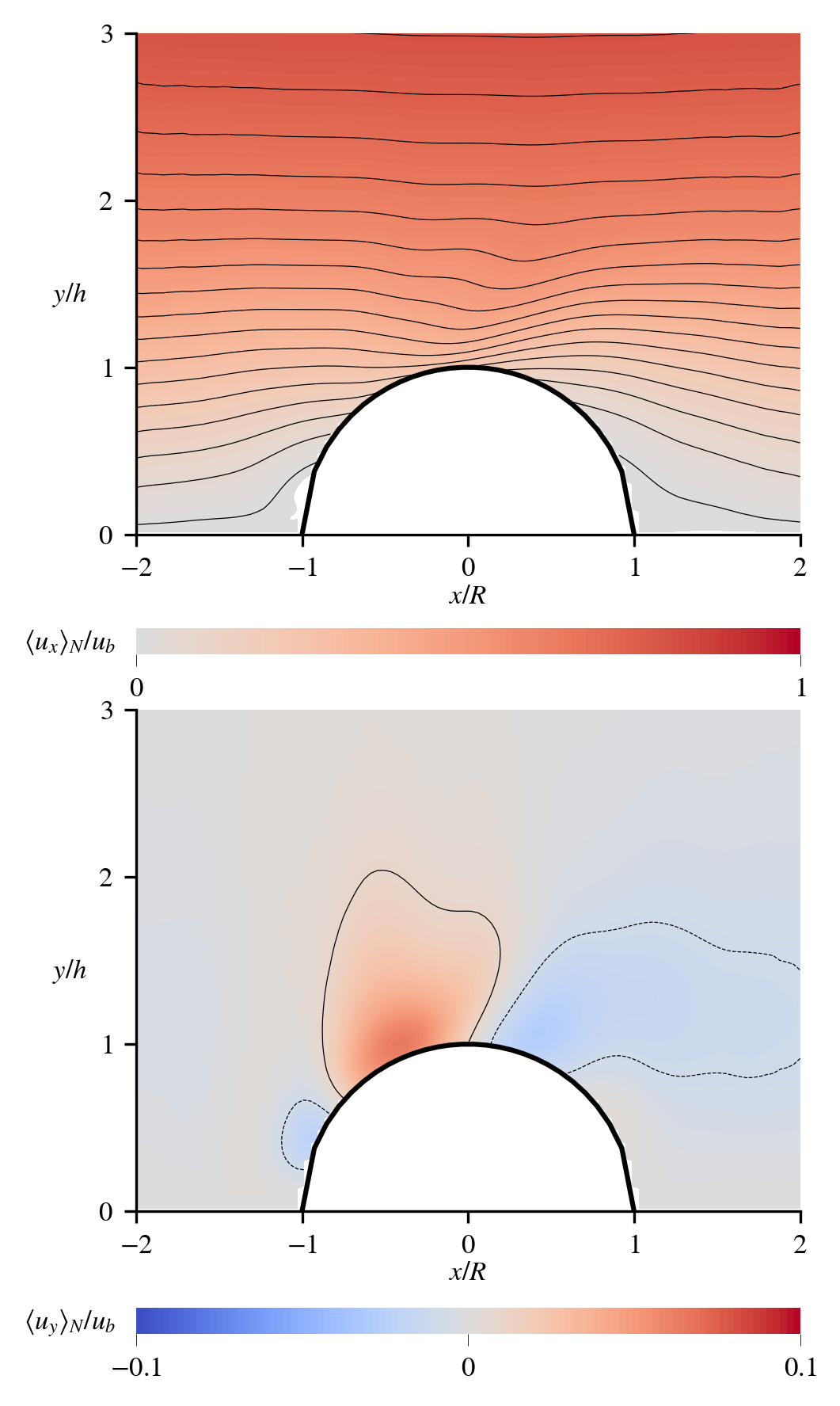}
\caption{The ensemble-averaged streamwise (top) and wall-normal velocity (bottom) around the droplet, normalized with the bulk velocity $u_b$. Visible is the strong deflection upwards just before the droplet crest as well as the streamwise  velocity deficit in the droplet wake. Contours spaced by $0.05$ for $\langle u_x\rangle_N$, and at $\pm 0.01$ for $\langle u_y\rangle_N$. \label{fig:droplet_ux_uy}}
\end{figure}

Figure \ref{fig:droplet_ux_uy} shows the ensemble averages of the streamwise ($\langle u_x\rangle_N$) and wall-normal  ($\langle u_y\rangle_N$) velocity components in the $x-y$ plane centered on the droplet.
Deflection of the flow around the droplet is clearly visible.
Fluid impinges on the upstream flank of the droplet, with a strong updraft induced near the droplet crest.
The recirculation found in the instantaneous velocity behind droplets near the inlet does not persist in the combined average over time and all droplets.
The influence of the droplet extends into the volume, with modifications of the flow contours visible beyond twice the normalized height $y/h$.
Measured at the downstream edge of the droplet footprint, the integrated deficit up to $y^+=15$ in the average streamwise flow velocity is $2.3\%$, with a peak deficit of $6.4\%$ at a height of $y=1.06h$.

\begin{figure}[h]
\centering
\includegraphics[scale=1]{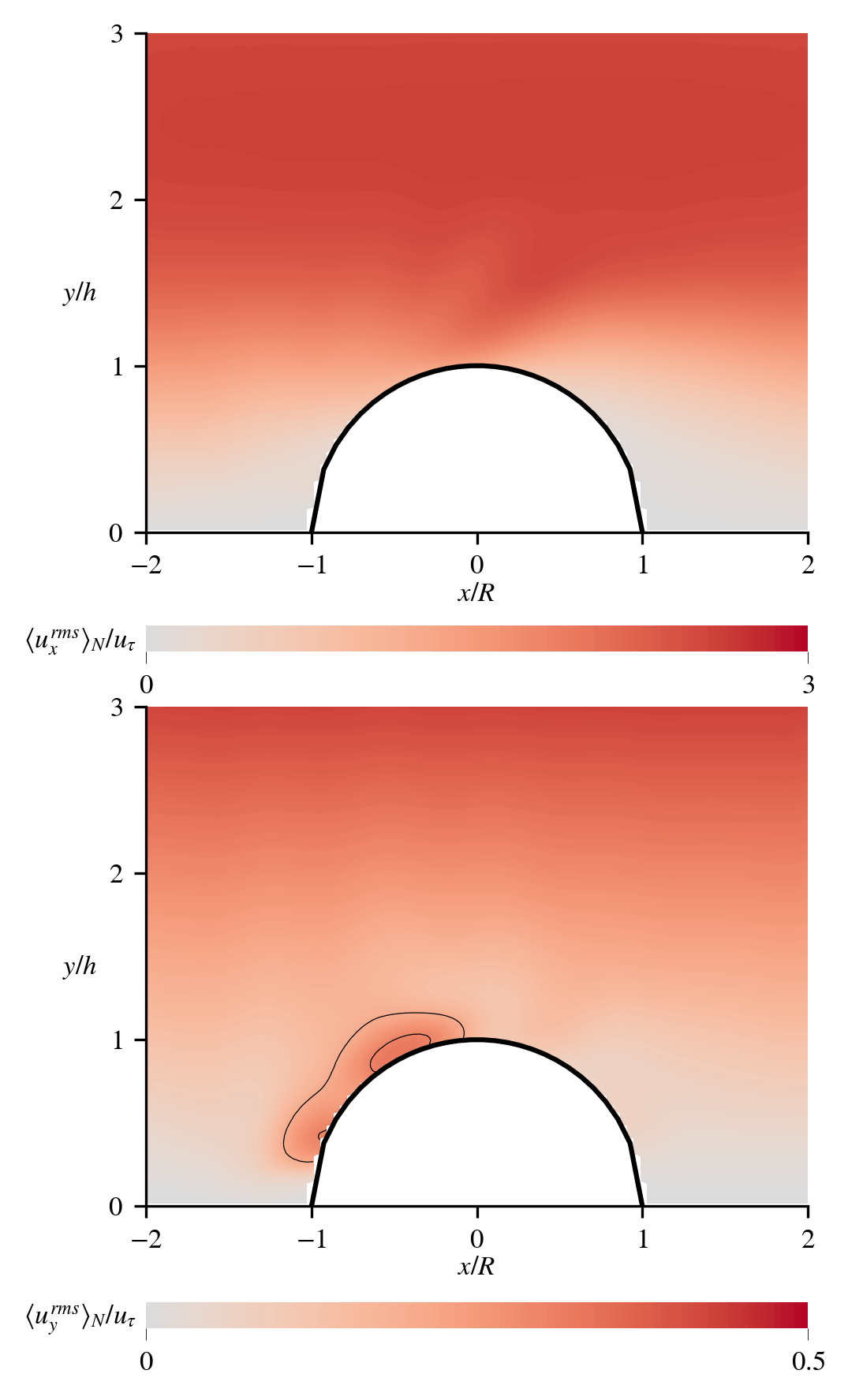}
\caption{The ensemble-averaged fluctuations of the streamwise (top) and wall-normal velocity components (bottom) around the droplet, normalized with the friction velocity $u_\tau$. Of particular note is the region of strong $u_y$-fluctuations on the upstream flank of the droplet. \label{fig:droplet_uxrms_uyrms}}
\end{figure}

Alongside the mean velocity field, the ensemble-averaged root-mean-square (rms) fluctuations of the streamwise ($\langle u_x^{\mathit{rms}}\rangle_N$) and wall-normal ($\langle u_y^{\mathit{rms}}\rangle_N$) velocity components in Figure \ref{fig:droplet_uxrms_uyrms} show strong fluctuations in the wall-normal component at upper part of the upstream flank of the droplet, where the deformation starts to reach out of the viscous sublayer and interacts with the turbulent flow through the channel.
In contrast, fluctuations of both the streamwise and wall-normal velocity are reduced in the droplet wake, where the obstruction by the droplet provides a degree of isolation of downstream region from the flow.
This is consistent with the observation of the deficit in the mean velocity discussed above.

\begin{figure}[h]
\centering
\includegraphics[scale=1]{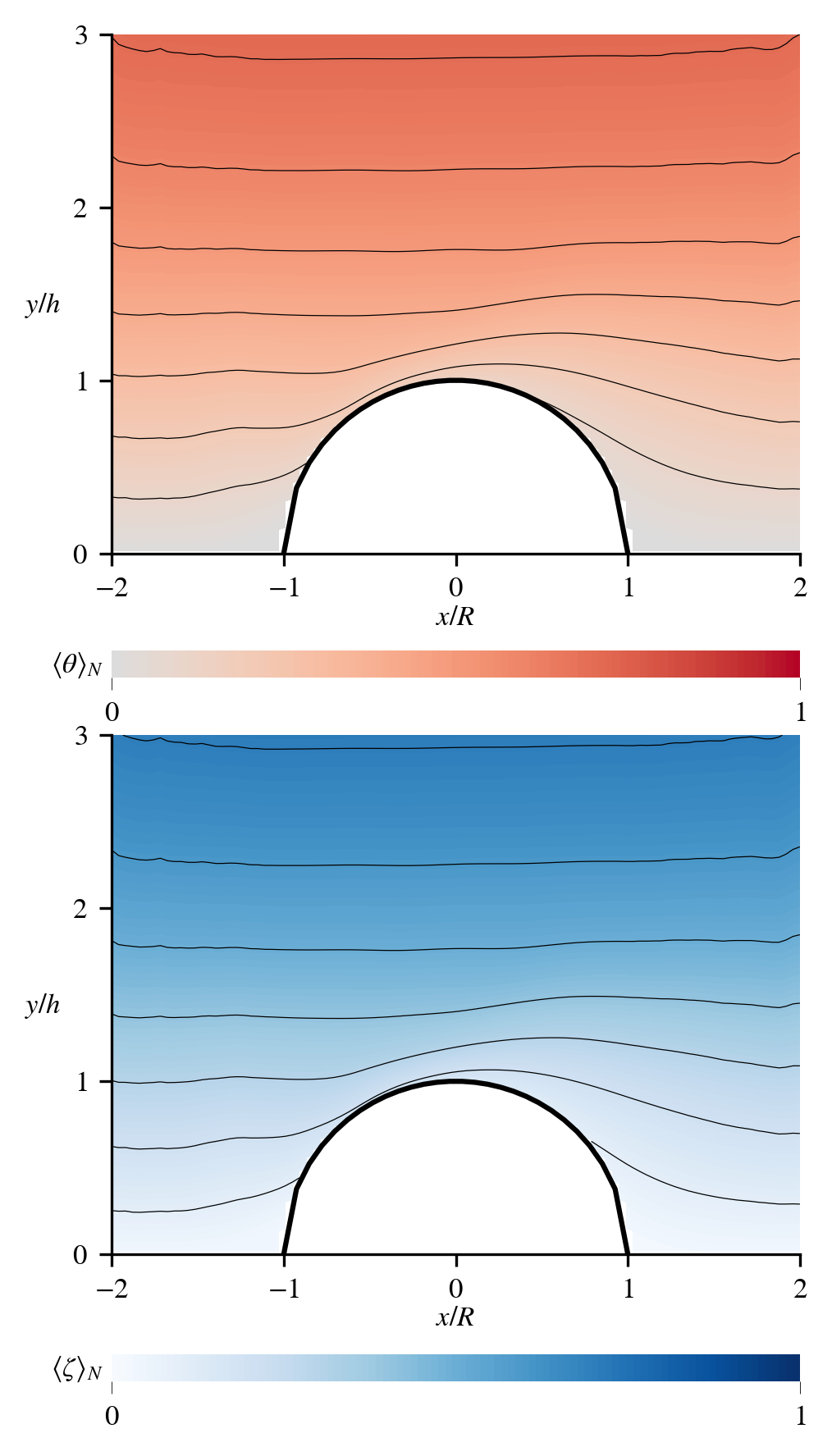}
\caption{The ensemble-averaged dimensionless temperature (top) and vapor concentration (bottom) around the droplet. The deformation of the field above the droplets and depletion of heat and humidity in the droplet wake are visible in the contours, spaced in intervals of $0.1$. \label{fig:droplet_theta_zeta}}
\end{figure}

The changes to the convective transport of heat and vapor caused by the deflection of the flow around the droplets and the interaction with the turbulence in turn affect the temperature and vapor concentration fields around the droplets.
Shown in Figure \ref{fig:droplet_theta_zeta} are the ensemble-averaged dimensionless temperature and vapor, with
\begin{align}
  \theta = \frac{T - T_c}{\Delta T},\label{eq:theta_definition}
\end{align}
defined analogous to the dimensionless vapor concentration in eq. (\ref{eq:zeta_definition}).
In the droplet wake, warm and humid fluid is depleted compared to the upstream side, mirroring the deficit in the streamwise velocity.
The cooling and drying of the fluid at the droplet surface, combined with the deflection of the flow around the droplet, impede the transport of heat and vapor from the bulk into the wake region.
As was the case for the velocity fields, the influence of the droplet on the scalar fields extends beyond the height of the deformation and is still visible up to $y\simeq 2h$.
Steep temperature and humidity gradients at the upstream flank up to the droplet crest coincide with the location of strong fluctuations in the wall-normal velocity seen in Figure \ref{fig:droplet_uxrms_uyrms}.
Increased turbulent mixing in this region causes strong transport of vapor and heat towards the droplet surface.

\begin{figure}[h]
\centering
\includegraphics[scale=1]{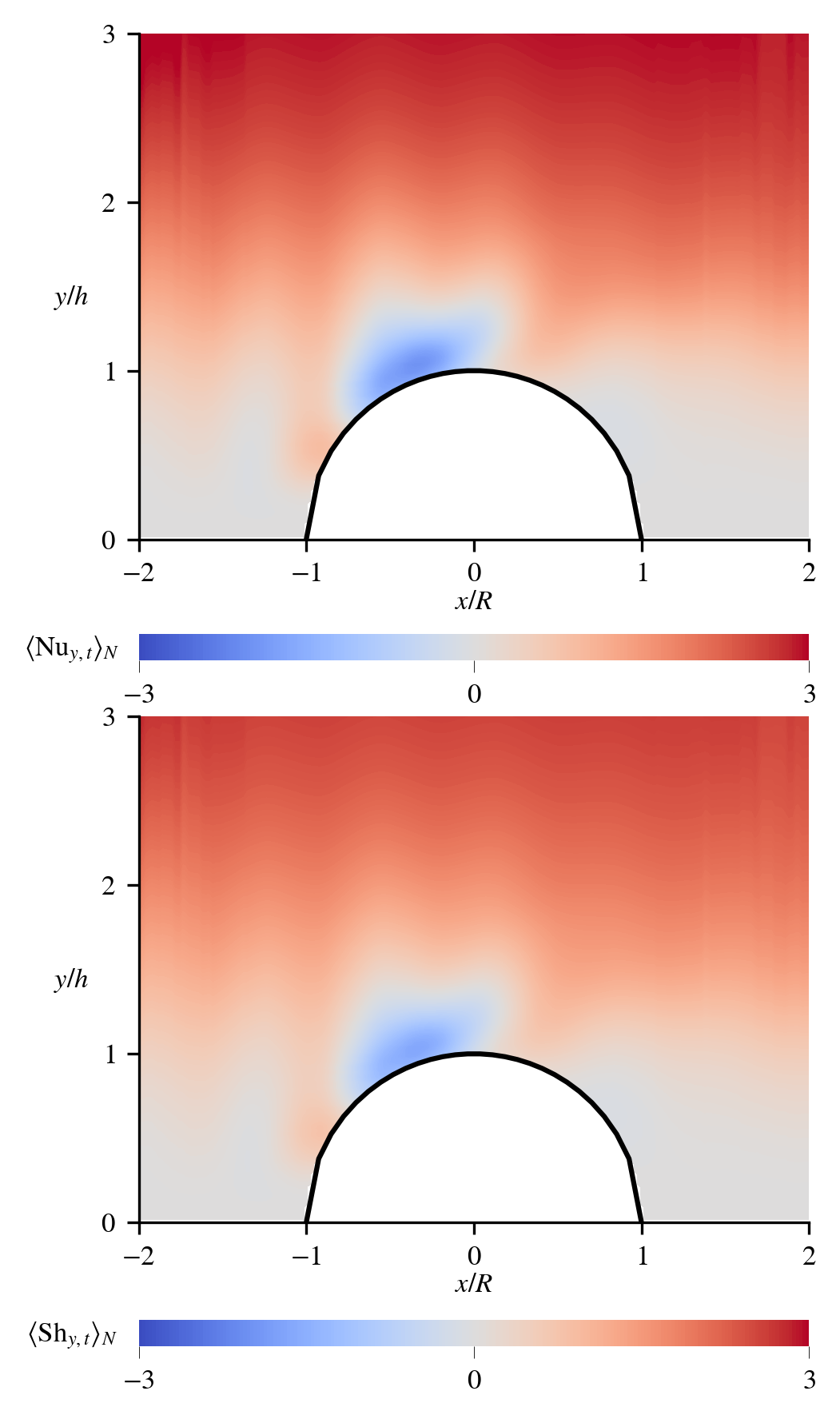}
\caption{The wall-normal turbulent transport as quantified by the ensemble-averaged turbulent Nusselt number for the heat transport (top) and Sherwood number for vapor transport (bottom). Both quantities behave similarly, again showing strong transport processes directly at the upstream flank of the droplet. }\label{fig:droplet_Nu_Sh}
\end{figure}

For the influence of the droplets on further condensation, the turbulent transport of the scalar fields is of particular importance.
Showing the ensemble average of the turbulent Nusselt number in the wall-normal direction,\cite{Kuerten2011}
\begin{align}
  \mathit{Nu}_{t,y} = -\frac{\delta}{\kappa}\,\left\langle u^\prime_y \theta^\prime\right\rangle,\label{eq:Nusselt_number}
\end{align}
with the fluctuations of the velocity $u^\prime$ and temperature $\theta^\prime$ as well as the similarly defined Sherwood number,
\begin{align}
  \mathit{Sh}_{t,y} = -\frac{\delta}{D}\,\left\langle u^\prime_y \zeta^\prime\right\rangle,\label{eq:Sherwood_number}
\end{align}
in Figure \ref{fig:droplet_Nu_Sh} illustrates the modification of the turbulent transport due to the droplets.
Since the Prandtl and Schmidt numbers are similar, and both temperature and concentration experience the same convection, the distribution of Nusselt and Sherwood number around the droplets mirror each other.
Again, the region directly at the upstream flank of the droplet exhibits the largest effect, showing a strong correlation between movement of the fluid towards the droplet and negative concentration fluctuations due to condensation at the surface.
As warm and humid fluid originating from the bulk flows towards the droplet in turbulent sweeps, it is loses heat in the interaction with the surface, and, as the temperature drops below the dew point, simultaneously transfers vapor across the phase boundary during condensation.
Critically, far less turbulent heat and vapor transfer occur in the region surrounding the droplet.

\begin{figure}[h]
\centering
\includegraphics[scale=1]{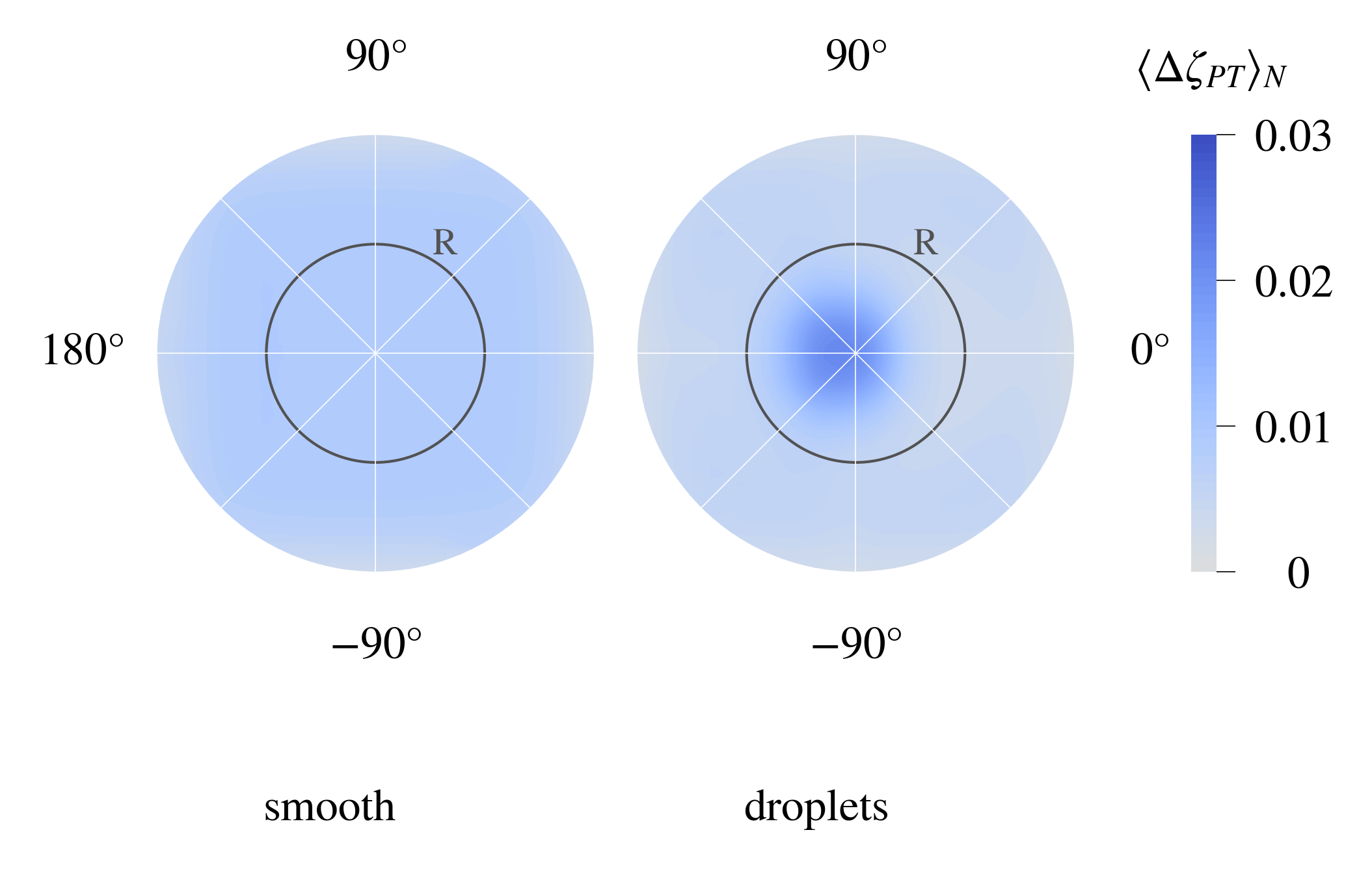}
\caption{The distribution of the ensemble-averaged condensation rate across the droplet (enclosed by its perimeter with radius $R$) and surrounding area, shown on the right, compared to the equivalent region in the simulation with a smooth wall on the left.} \label{fig:mdot_polar}
\end{figure}

This increased vapor transport rate directly affects the condensation rate at the cooled wall.
Figure \ref{fig:mdot_polar} shows a top-down view of ensemble-averaged dimensionless condensation rate $\langle \Delta \zeta_\mathit{PT}\rangle_N$ at the wall, compared between the baseline simulation without droplets on the left and with the deformed wall on the right.
The overall non-homogeneity of the condensation rate across the complete channel wall notwithstanding, condensation rates are uniform on the smooth wall on length scales comparable the droplet radius.
In the presence of condensate droplets, however, the symmetry of the cooled wall is broken, and condensation is focussed on the surface of the droplets, shifted from the surrounding smooth surface, where condensation rates are reduced in comparison.
In fact, the region exhibiting the highest condensation rates is the upstream flank of the droplet, extending up onto the crest.
This corresponds exactly to the region of increased wall-normal velocity fluctuations and increased vapor transport discussed above in Figure \ref{fig:droplet_Nu_Sh}, confirming the conclusion that here, scalar transport towards the droplet is markedly enhanced by the influence of the wall deformation on the turbulent flow.

\begin{figure}[h]
\centering
\includegraphics[scale=1]{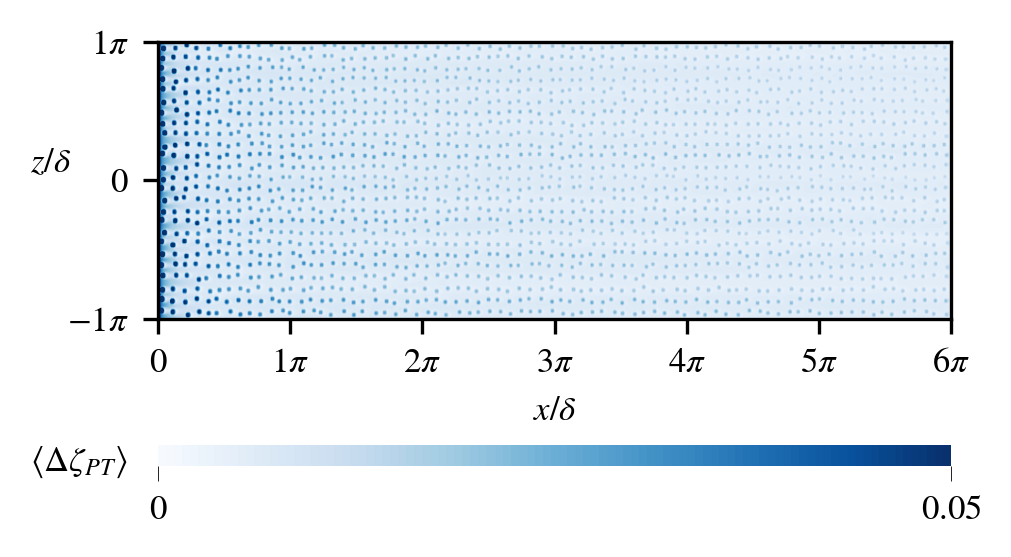}
\caption{The distribution of the average condensation rate at the cooled wall with sessile condensate droplets. Condensation is strongly focussed on the droplets, while the streamwise and spanwise variability from the smooth case persists.\label{fig:mdot_droplets}}
\end{figure}

Across the complete channel wall, the behavior observed in the average over the droplet ensemble results in a markedly different distribution of the condensation rate compared to the smooth wall in Figure \ref{fig:mdot_smooth}.
Figure \ref{fig:mdot_droplets} shows the result for the simulation with droplets at the wall.
The existence of the surface deformations breaks the symmetry of the wall, introducing regions which strongly favor condensation above the surrounding areas.
On top of the inhomogeneity introduced by the droplets, the variations of the condensate rate along the channel observed with the smooth wall, with higher rates near the inlet and variations across the spanwise direction are still present.
They result from features of the overall flow, with the diminishing available vapor further along the channel and the spanwise structures found in the flow unaffected by the surface modifications.\cite{Grass1971}

Condensation preferentially occurring on the surface of existing droplets mirrors the thermodynamic phenomenon of Ostwald ripening,\cite{Madras2001} where larger droplets experience more additional condensation than smaller droplets due to the difference in surface energy of the curved interface.
Since this thermodynamic effect is not included in the simulation, observing similar behavior reveals an analogous effect stemming from the interaction between turbulent flow and the droplet geometry alone.

\section{Global heat and mass transfer modifications}
The simulation approach focussing only on the gas phase and modeling the condensate droplets as static wall deformations allows the investigation of the impact of condensation across the full length of the channel.
Following from the effect of the droplets on the transport processes at the wall, the macroscopic evolution of both scalar fields is affected by the added condensate droplets.
Considering both temperature and vapor concentration in terms of the associated thermal energy flux along the streamwise direction allows the combined discussion of both quantities.
The sensitive heat flux is
\begin{align}
  \phi_s = \langle\theta\rangle_b \Delta T \rho c_p u_b,\label{eq:sensitive_flux}
\end{align}
with the bulk-averaged dimensionless temperature
\begin{align}
  \langle \theta\rangle_b = \frac{\int \theta \vect{u}\cdot\mathrm{d}\vect{A}}{\int \vect{u}\cdot\mathrm{d}\vect{A}},\label{eq:bulk_average_theta}
\end{align}
where $A$ is the cross-sectional area of the channel.
Similarly, the latent heat flux is then given by
\begin{align}
  \phi_l = \langle\zeta\rangle_b \Delta c \rho_v h u_b,\label{eq:latent_flux}
\end{align}
with the bulk-averaged dimensionless concentration defined analogously to eq. (\ref{eq:bulk_average_theta}).

\begin{figure}[h]
\includegraphics[scale=1]{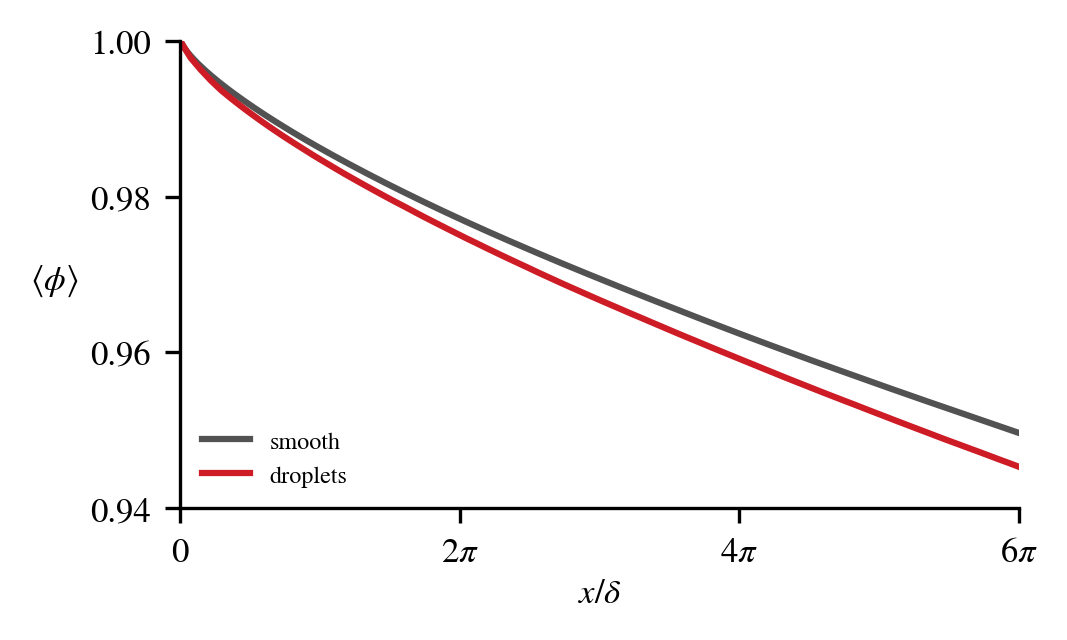}
\caption{Evolution of the normalized total amount of thermal energy flux along the channel. The presence of condensate droplets significantly increases the energy transfer at the cooled wall, leading to a larger deficit at the outlet.\label{fig:energy_flux}}
\end{figure}

Adding both contributions and normalizing with the total flux at the channel inlet $\phi_{s}^\mathit{in}+\phi_{l}^\mathit{in}$ gives
\begin{align}
  \phi = \frac{\phi_s + \phi_l}{\phi_s^{\mathit{in}} + \phi_l^{\mathit{in}}}=\frac{\langle\theta\rangle_b + (\tilde{\rho}/\mathit{Ja})\,\langle\zeta\rangle_b}{1+ (\tilde{\rho}/\mathit{Ja})},\label{eq:normalized_flux}
\end{align}
the total remaining heat flux at position $x$ along the channel, with the density ratio $\tilde{\rho}=\Delta c \rho_v/\rho$ between the effective vapor density and the density of the fluid as a whole.
Figure \ref{fig:energy_flux} shows the progression of $\phi$ compared for the simulations with and without droplets.
The deficit in heat flux through the cross section of the channel compared to the inlet is equal to the accumulated flux through the cooled wall.
Along the whole channel, more thermal energy is removed through the droplet-covered wall compared to the smooth wall.
At the outlet, the accumulated deficit amounts to $\Delta \phi=0.050$ without and $\Delta \phi=0.055$ with droplets, for an increase of $10\%$.
Critically, this is larger than the pure increase of the wall surface by $6.4\%$ from the addition of the droplets, confirming additional energy transfer due to changes in the turbulent flow.

\begin{figure}[h]
\includegraphics[scale=1]{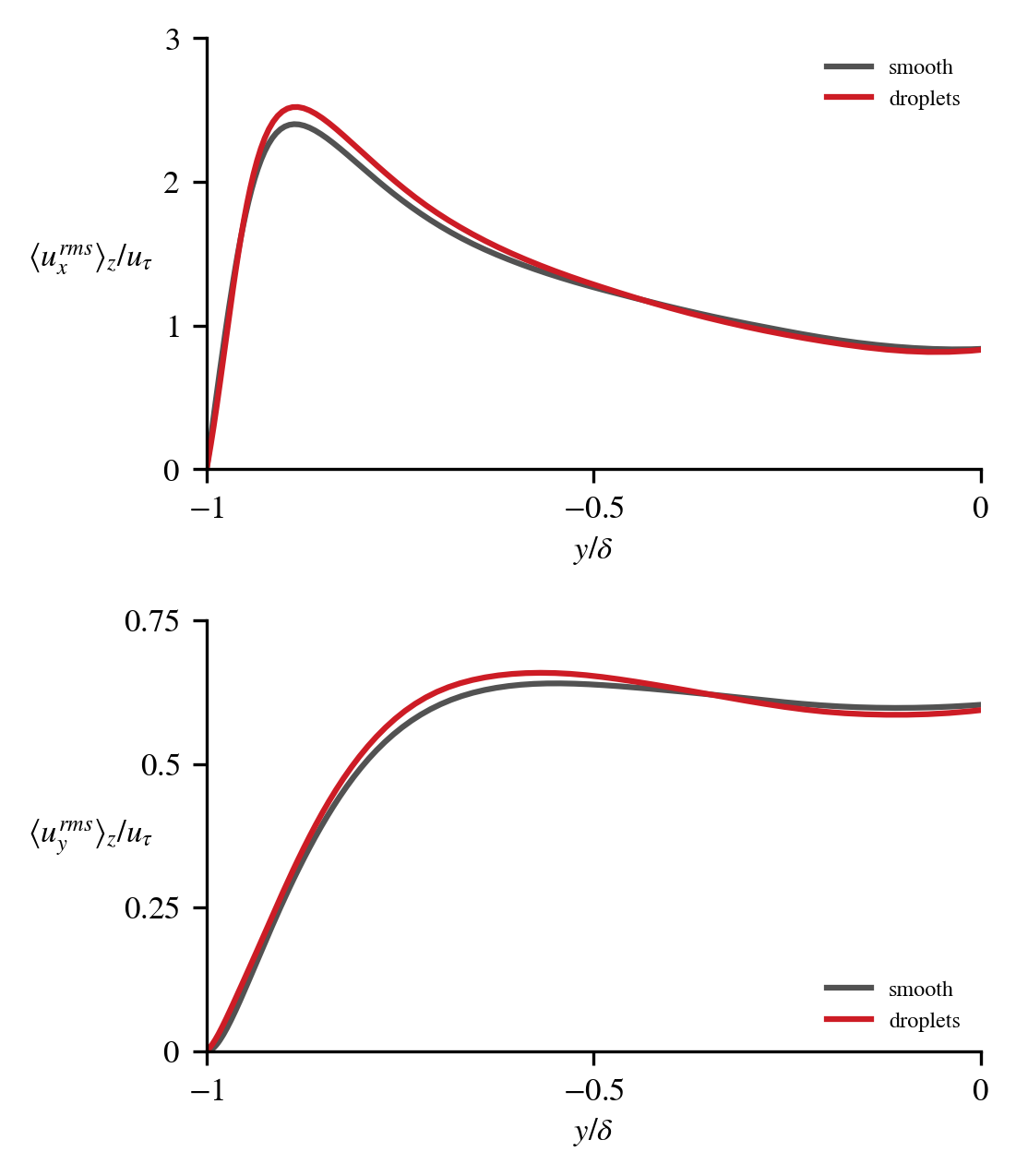}
\caption{Fluctuations of the streamwise (top) and wall-normal velocity, compared between simulations with and without droplets. \label{fig:urms}}
\end{figure}

To balance the additional thermal energy removed from the channel at the wall in the presence of droplets, the transport of heat and vapor inside the channel towards the wall must be increased accordingly.
The modification of the turbulent transport on the scale of the complete channel is the continuation of the modified flow fields in the immediate vicinity of the droplets.
In addition to temporal averaging, quantities are averaged over the spanwise coordinate $z$, denoted by $\langle\cdot\rangle_z$
Figure \ref{fig:urms} shows wall-normal profiles of the rms-fluctuations of the streamwise and wall-normal components of the velocity at the outlet of the channel.
At the channel outlet, the accumulated effects from condensation and the droplets are largest, as the flow was exposed to their influence over the complete residence time of the fluid in the channel.
Both velocity components show increased fluctuations in the case with condensate droplets caused by the deflection of the flow around the individual droplets, as shown above in Figures \ref{fig:droplet_ux_uy} and \ref{fig:droplet_uxrms_uyrms}.
This combines with the opposing buoyancy resulting from the release of latent heat during condensation, with acts against the aiding buoyant force due to the cooling at the wall.
The overall reduction of the effectiveness of the aiding buoyancy as a consequence of condensation limits its damping effect,\cite{Kasagi1997,Wetzel2019} thereby increasing the turbulence near the cooled wall.

\begin{figure}[h]
\includegraphics[scale=1]{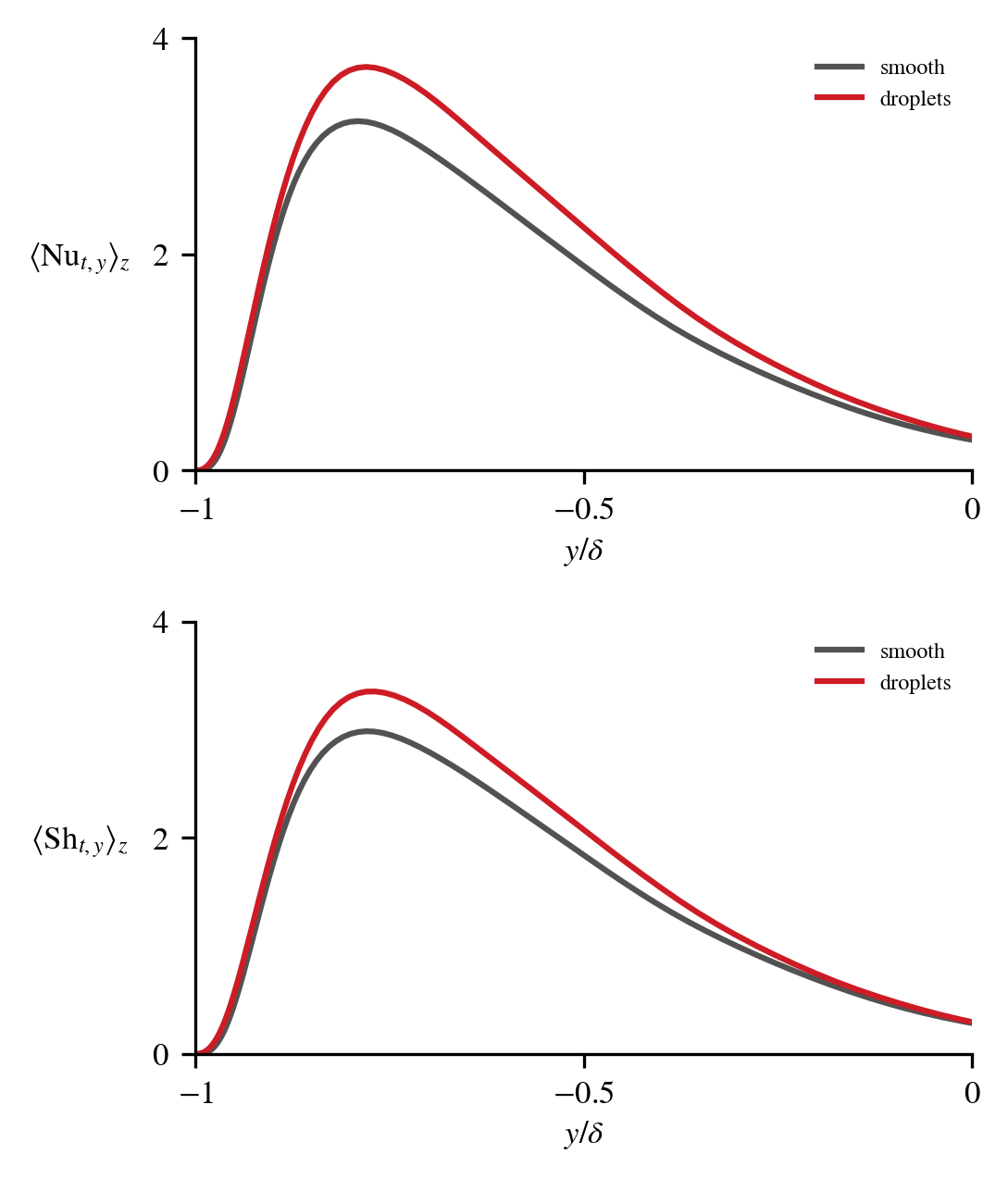}
\caption{Comparison of the turbulent Nusselt (top) and Sherwood numbers (bottom) with and without droplets at the cooled wall, averaged in time and along the spanwise direction. \label{fig:Nusselt_Sherwood}}
\end{figure}

Increased mixing from the changes to the velocity field then impacts the turbulent transport of heat and vapor in the wall-normal direction.
Higher turbulence intensity leads to more transfer of warm and humid fluid in sweeps, moving from the bulk towards the cooled wall.
Figure \ref{fig:Nusselt_Sherwood} shows the marked change of the dimensionless heat and mass transfer in the form of the Nusselt and Sherwood numbers in the simulation including droplets, with peaks increasing by $15.8\%$ and $12.4\%$, respectively.

\section{Conclusions}
For the range of temperatures and humidity typically encountered at ambient conditions and therefore relevant for ventilation applications, condensation on cooled surfaces dominates compared to droplet formation within the flow volume.
In these circumstances, the combination of disregarding the dispersed liquid phase within the volume and modeling the sessile condensate droplets at the wall as static deformations allows DNS of flows with phase transition while incurring none of the computational costs associated with other multiphase DNS techniques.

Because of the slow growth dynamics of droplets at the condensation rates found at ambient conditions, droplets act as static obstacles with respect to the turbulent flow.
While the separation of time scales between flow and droplets makes fully resolved simulations of the growth of droplet patterns infeasible, the single-phase approach utilizes the average condensation rates obtained over a time interval on the scale of the convective time scale to extrapolate the accumulation of condensate mass over the much longer time scales of droplet growth, thereby using the large difference between both scales to its advantage.

Investigating the effects of the droplet pattern on the flow through the channel confirms that large sessile condensate droplets on the wall surface are an important aspect of flows with condensation.
Their presence significantly alters the flow in the sensitive wall-near region, and the interactions with the turbulent flow project the influence of the droplets outward towards the bulk.
In particular, the increase of turbulent mixing leads to an increase in the turbulent transport of heat and mass towards the cooled surface.
Here, the breaking of the uniformity of the wall concentrates further condensation on the surface of existing droplets in a positive feed-back effect.

Including static, sessile condensate droplets extends the previously established single-phase approach for DNS of flows with surface condensation to include the significant contributions of the condensate to the overall effect of condensation on the flow and the transport of heat and mass.
The flexibility of the approach as presented here allows for refinement to include additional aspects of dropwise condensation, such as constraining the droplet pattern to not only conserve mass locally, but respect a global size distribution, or the modification of the saturation pressure at curved interfaces, without compromising the computational cost advantages inherent in the method.

\bibliography{2022journal}

\begin{thebibliography}{30}%
\makeatletter
\providecommand \@ifxundefined [1]{%
 \@ifx{#1\undefined}
}%
\providecommand \@ifnum [1]{%
 \ifnum #1\expandafter \@firstoftwo
 \else \expandafter \@secondoftwo
 \fi
}%
\providecommand \@ifx [1]{%
 \ifx #1\expandafter \@firstoftwo
 \else \expandafter \@secondoftwo
 \fi
}%
\providecommand \natexlab [1]{#1}%
\providecommand \enquote  [1]{``#1''}%
\providecommand \bibnamefont  [1]{#1}%
\providecommand \bibfnamefont [1]{#1}%
\providecommand \citenamefont [1]{#1}%
\providecommand \href@noop [0]{\@secondoftwo}%
\providecommand \href [0]{\begingroup \@sanitize@url \@href}%
\providecommand \@href[1]{\@@startlink{#1}\@@href}%
\providecommand \@@href[1]{\endgroup#1\@@endlink}%
\providecommand \@sanitize@url [0]{\catcode `\\12\catcode `\$12\catcode
  `\&12\catcode `\#12\catcode `\^12\catcode `\_12\catcode `\%12\relax}%
\providecommand \@@startlink[1]{}%
\providecommand \@@endlink[0]{}%
\providecommand \url  [0]{\begingroup\@sanitize@url \@url }%
\providecommand \@url [1]{\endgroup\@href {#1}{\urlprefix }}%
\providecommand \urlprefix  [0]{URL }%
\providecommand \Eprint [0]{\href }%
\providecommand \doibase [0]{http://dx.doi.org/}%
\providecommand \selectlanguage [0]{\@gobble}%
\providecommand \bibinfo  [0]{\@secondoftwo}%
\providecommand \bibfield  [0]{\@secondoftwo}%
\providecommand \translation [1]{[#1]}%
\providecommand \BibitemOpen [0]{}%
\providecommand \bibitemStop [0]{}%
\providecommand \bibitemNoStop [0]{.\EOS\space}%
\providecommand \EOS [0]{\spacefactor3000\relax}%
\providecommand \BibitemShut  [1]{\csname bibitem#1\endcsname}%
\let\auto@bib@innerbib\@empty
\bibitem [{\citenamefont {Lorenz}(2015)}]{Lorenz2015}%
  \BibitemOpen
  \bibfield  {author} {\bibinfo {author} {\bibfnamefont {M.}~\bibnamefont
  {Lorenz}},\ }\emph {\bibinfo {title} {Reduction of heating loads and interior
  window fogging in vehicles}},\ \href@noop {} {Ph.D. thesis},\ \bibinfo
  {school} {Technische Universität München} (\bibinfo {year}
  {2015})\BibitemShut {NoStop}%
\bibitem [{\citenamefont {Leriche}\ \emph {et~al.}(2015)\citenamefont
  {Leriche}, \citenamefont {Roessner}, \citenamefont {Reister},\ and\
  \citenamefont {Weigand}}]{Leriche2015}%
  \BibitemOpen
  \bibfield  {author} {\bibinfo {author} {\bibfnamefont {M.}~\bibnamefont
  {Leriche}}, \bibinfo {author} {\bibfnamefont {W.}~\bibnamefont {Roessner}},
  \bibinfo {author} {\bibfnamefont {H.}~\bibnamefont {Reister}}, \ and\
  \bibinfo {author} {\bibfnamefont {B.}~\bibnamefont {Weigand}},\ }\bibfield
  {title} {\enquote {\bibinfo {title} {Numerical investigation of droplets
  condensation on a windshield: Prediction of fogging behavior},}\ }in\ \href
  {\doibase 10.4271/2015-01-0360} {\emph {\bibinfo {booktitle} {{SAE} Technical
  Paper Series}}}\ (\bibinfo  {publisher} {{SAE} International},\ \bibinfo
  {year} {2015})\BibitemShut {NoStop}%
\bibitem [{\citenamefont {Balachandar}\ and\ \citenamefont
  {Eaton}(2010)}]{Balachandar2010}%
  \BibitemOpen
  \bibfield  {author} {\bibinfo {author} {\bibfnamefont {S.}~\bibnamefont
  {Balachandar}}\ and\ \bibinfo {author} {\bibfnamefont {J.~K.}\ \bibnamefont
  {Eaton}},\ }\bibfield  {title} {\enquote {\bibinfo {title} {Turbulent
  dispersed multiphase flow},}\ }\href {\doibase
  10.1146/annurev.fluid.010908.165243} {\bibfield  {journal} {\bibinfo
  {journal} {Annual Review of Fluid Mechanics}\ }\textbf {\bibinfo {volume}
  {42}},\ \bibinfo {pages} {111--133} (\bibinfo {year} {2010})}\BibitemShut
  {NoStop}%
\bibitem [{\citenamefont {Kuerten}(2016)}]{Kuerten2016}%
  \BibitemOpen
  \bibfield  {author} {\bibinfo {author} {\bibfnamefont {J.~G.~M.}\
  \bibnamefont {Kuerten}},\ }\bibfield  {title} {\enquote {\bibinfo {title}
  {Point-particle {DNS} and {LES} of particle-laden turbulent flow - a
  state-of-the-art review},}\ }\href {\doibase 10.1007/s10494-016-9765-y}
  {\bibfield  {journal} {\bibinfo  {journal} {Flow, Turbulence and Combustion}\
  }\textbf {\bibinfo {volume} {97}},\ \bibinfo {pages} {689--713} (\bibinfo
  {year} {2016})}\BibitemShut {NoStop}%
\bibitem [{\citenamefont {Russo}\ \emph {et~al.}(2014)\citenamefont {Russo},
  \citenamefont {Kuerten}, \citenamefont {van~der Geld},\ and\ \citenamefont
  {Geurts}}]{Russo2014}%
  \BibitemOpen
  \bibfield  {author} {\bibinfo {author} {\bibfnamefont {E.}~\bibnamefont
  {Russo}}, \bibinfo {author} {\bibfnamefont {J.~G.~M.}\ \bibnamefont
  {Kuerten}}, \bibinfo {author} {\bibfnamefont {C.~W.~M.}\ \bibnamefont
  {van~der Geld}}, \ and\ \bibinfo {author} {\bibfnamefont {B.~J.}\
  \bibnamefont {Geurts}},\ }\bibfield  {title} {\enquote {\bibinfo {title}
  {Water droplet condensation and evaporation in turbulent channel flow},}\
  }\href {\doibase 10.1017/jfm.2014.239} {\bibfield  {journal} {\bibinfo
  {journal} {Journal of Fluid Mechanics}\ }\textbf {\bibinfo {volume} {749}},\
  \bibinfo {pages} {666--700} (\bibinfo {year} {2014})}\BibitemShut {NoStop}%
\bibitem [{\citenamefont {Orazzo}\ and\ \citenamefont
  {Tanguy}(2019)}]{Orazzo2019}%
  \BibitemOpen
  \bibfield  {author} {\bibinfo {author} {\bibfnamefont {A.}~\bibnamefont
  {Orazzo}}\ and\ \bibinfo {author} {\bibfnamefont {S.}~\bibnamefont
  {Tanguy}},\ }\bibfield  {title} {\enquote {\bibinfo {title} {Direct numerical
  simulations of droplet condensation},}\ }\href {\doibase
  10.1016/j.ijheatmasstransfer.2018.07.094} {\bibfield  {journal} {\bibinfo
  {journal} {International Journal of Heat and Mass Transfer}\ }\textbf
  {\bibinfo {volume} {129}},\ \bibinfo {pages} {432--448} (\bibinfo {year}
  {2019})}\BibitemShut {NoStop}%
\bibitem [{\citenamefont {Bahavar}\ and\ \citenamefont
  {Wagner}(2020)}]{Bahavar2020}%
  \BibitemOpen
  \bibfield  {author} {\bibinfo {author} {\bibfnamefont {P.}~\bibnamefont
  {Bahavar}}\ and\ \bibinfo {author} {\bibfnamefont {C.}~\bibnamefont
  {Wagner}},\ }\bibfield  {title} {\enquote {\bibinfo {title}
  {Condensation-induced flow structure modifications in turbulent channel flow
  investigated in direct numerical simulations},}\ }\href {\doibase
  10.1063/1.5128976} {\bibfield  {journal} {\bibinfo  {journal} {Physics of
  Fluids}\ }\textbf {\bibinfo {volume} {32}},\ \bibinfo {pages} {015115}
  (\bibinfo {year} {2020})}\BibitemShut {NoStop}%
\bibitem [{\citenamefont {Jiménez}\ and\ \citenamefont
  {Moin}(1991)}]{Jimenez1991}%
  \BibitemOpen
  \bibfield  {author} {\bibinfo {author} {\bibfnamefont {J.}~\bibnamefont
  {Jiménez}}\ and\ \bibinfo {author} {\bibfnamefont {P.}~\bibnamefont
  {Moin}},\ }\bibfield  {title} {\enquote {\bibinfo {title} {The minimal flow
  unit in near-wall turbulence},}\ }\href {\doibase 10.1017/S0022112091002033}
  {\bibfield  {journal} {\bibinfo  {journal} {Journal of Fluid Mechanics}\
  }\textbf {\bibinfo {volume} {225}},\ \bibinfo {pages} {213–240} (\bibinfo
  {year} {1991})}\BibitemShut {NoStop}%
\bibitem [{\citenamefont {Hammou}\ \emph {et~al.}(2004)\citenamefont {Hammou},
  \citenamefont {Benhamou}, \citenamefont {Galanis},\ and\ \citenamefont
  {Orfi}}]{Hammou2004}%
  \BibitemOpen
  \bibfield  {author} {\bibinfo {author} {\bibfnamefont {Z.~A.}\ \bibnamefont
  {Hammou}}, \bibinfo {author} {\bibfnamefont {B.}~\bibnamefont {Benhamou}},
  \bibinfo {author} {\bibfnamefont {N.}~\bibnamefont {Galanis}}, \ and\
  \bibinfo {author} {\bibfnamefont {J.}~\bibnamefont {Orfi}},\ }\bibfield
  {title} {\enquote {\bibinfo {title} {Laminar mixed convection of humid air in
  a vertical channel with evaporation or condensation at the wall},}\ }\href
  {\doibase 10.1016/j.ijthermalsci.2003.10.010} {\bibfield  {journal} {\bibinfo
   {journal} {International Journal of Thermal Sciences}\ }\textbf {\bibinfo
  {volume} {43}},\ \bibinfo {pages} {531--539} (\bibinfo {year}
  {2004})}\BibitemShut {NoStop}%
\bibitem [{\citenamefont {Gray}\ and\ \citenamefont
  {Giorgini}(1976)}]{Gray1976}%
  \BibitemOpen
  \bibfield  {author} {\bibinfo {author} {\bibfnamefont {D.~D.}\ \bibnamefont
  {Gray}}\ and\ \bibinfo {author} {\bibfnamefont {A.}~\bibnamefont
  {Giorgini}},\ }\bibfield  {title} {\enquote {\bibinfo {title} {The validity
  of the boussinesq approximation for liquids and gases},}\ }\href {\doibase
  10.1016/0017-9310(76)90168-X} {\bibfield  {journal} {\bibinfo  {journal}
  {International Journal of Heat and Mass Transfer}\ }\textbf {\bibinfo
  {volume} {19}},\ \bibinfo {pages} {545 -- 551} (\bibinfo {year}
  {1976})}\BibitemShut {NoStop}%
\bibitem [{\citenamefont {Marek}\ and\ \citenamefont
  {Straub}(2001)}]{Marek2001}%
  \BibitemOpen
  \bibfield  {author} {\bibinfo {author} {\bibfnamefont {R.}~\bibnamefont
  {Marek}}\ and\ \bibinfo {author} {\bibfnamefont {J.}~\bibnamefont {Straub}},\
  }\bibfield  {title} {\enquote {\bibinfo {title} {Analysis of the evaporation
  coeffcient and the condensation coeffcient of water},}\ }\href {\doibase
  10.1016/S0017-9310(00)00086-7} {\bibfield  {journal} {\bibinfo  {journal}
  {International Journal of Heat and Mass Transfer}\ }\textbf {\bibinfo
  {volume} {44}},\ \bibinfo {pages} {39--53} (\bibinfo {year}
  {2001})}\BibitemShut {NoStop}%
\bibitem [{\citenamefont {Alduchov}\ and\ \citenamefont
  {Eskridge}(1996)}]{Alduchov1996}%
  \BibitemOpen
  \bibfield  {author} {\bibinfo {author} {\bibfnamefont {O.~A.}\ \bibnamefont
  {Alduchov}}\ and\ \bibinfo {author} {\bibfnamefont {R.~E.}\ \bibnamefont
  {Eskridge}},\ }\bibfield  {title} {\enquote {\bibinfo {title} {Improved
  magnus form approximation of saturation vapor pressure},}\ }\href {\doibase
  10.1175/1520-0450(1996)035<0601:imfaos>2.0.co;2} {\bibfield  {journal}
  {\bibinfo  {journal} {Journal of Applied Meteorology}\ }\textbf {\bibinfo
  {volume} {35}},\ \bibinfo {pages} {601--609} (\bibinfo {year}
  {1996})}\BibitemShut {NoStop}%
\bibitem [{\citenamefont {Tryggvason}, \citenamefont {Esmaeeli},\ and\
  \citenamefont {Al-Rawahi}(2005)}]{Tryggvason2005}%
  \BibitemOpen
  \bibfield  {author} {\bibinfo {author} {\bibfnamefont {G.}~\bibnamefont
  {Tryggvason}}, \bibinfo {author} {\bibfnamefont {A.}~\bibnamefont
  {Esmaeeli}}, \ and\ \bibinfo {author} {\bibfnamefont {N.}~\bibnamefont
  {Al-Rawahi}},\ }\bibfield  {title} {\enquote {\bibinfo {title} {Direct
  numerical simulations of flows with phase change},}\ }\href {\doibase
  10.1016/j.compstruc.2004.05.021} {\bibfield  {journal} {\bibinfo  {journal}
  {Computers {\&} Structures}\ }\textbf {\bibinfo {volume} {83}},\ \bibinfo
  {pages} {445--453} (\bibinfo {year} {2005})}\BibitemShut {NoStop}%
\bibitem [{\citenamefont {Niu}\ and\ \citenamefont {Tang}(2018)}]{Niu2018}%
  \BibitemOpen
  \bibfield  {author} {\bibinfo {author} {\bibfnamefont {D.}~\bibnamefont
  {Niu}}\ and\ \bibinfo {author} {\bibfnamefont {G.}~\bibnamefont {Tang}},\
  }\bibfield  {title} {\enquote {\bibinfo {title} {Molecular dynamics
  simulation of droplet nucleation and growth on a rough surface: revealing the
  microscopic mechanism of the flooding mode},}\ }\href {\doibase
  10.1039/c8ra04003f} {\bibfield  {journal} {\bibinfo  {journal} {{RSC}
  Advances}\ }\textbf {\bibinfo {volume} {8}},\ \bibinfo {pages} {24517--24524}
  (\bibinfo {year} {2018})}\BibitemShut {NoStop}%
\bibitem [{\citenamefont {Medici}\ \emph {et~al.}(2014)\citenamefont {Medici},
  \citenamefont {Mongruel}, \citenamefont {Royon},\ and\ \citenamefont
  {Beysens}}]{Medici2014}%
  \BibitemOpen
  \bibfield  {author} {\bibinfo {author} {\bibfnamefont {M.-G.}\ \bibnamefont
  {Medici}}, \bibinfo {author} {\bibfnamefont {A.}~\bibnamefont {Mongruel}},
  \bibinfo {author} {\bibfnamefont {L.}~\bibnamefont {Royon}}, \ and\ \bibinfo
  {author} {\bibfnamefont {D.}~\bibnamefont {Beysens}},\ }\bibfield  {title}
  {\enquote {\bibinfo {title} {Edge effects on water droplet condensation},}\
  }\href {\doibase 10.1103/PhysRevE.90.062403} {\bibfield  {journal} {\bibinfo
  {journal} {Phys. Rev. E}\ }\textbf {\bibinfo {volume} {90}},\ \bibinfo
  {pages} {062403} (\bibinfo {year} {2014})}\BibitemShut {NoStop}%
\bibitem [{\citenamefont {Kim}\ and\ \citenamefont {Kim}(2011)}]{Kim2011}%
  \BibitemOpen
  \bibfield  {author} {\bibinfo {author} {\bibfnamefont {S.}~\bibnamefont
  {Kim}}\ and\ \bibinfo {author} {\bibfnamefont {K.~J.}\ \bibnamefont {Kim}},\
  }\bibfield  {title} {\enquote {\bibinfo {title} {Dropwise condensation
  modeling suitable for superhydrophobic surfaces},}\ }\href {\doibase
  10.1115/1.4003742} {\bibfield  {journal} {\bibinfo  {journal} {Journal of
  Heat Transfer}\ }\textbf {\bibinfo {volume} {133}} (\bibinfo {year} {2011}),\
  10.1115/1.4003742}\BibitemShut {NoStop}%
\bibitem [{\citenamefont {Rose}(2002)}]{Rose2002}%
  \BibitemOpen
  \bibfield  {author} {\bibinfo {author} {\bibfnamefont {J.~W.}\ \bibnamefont
  {Rose}},\ }\bibfield  {title} {\enquote {\bibinfo {title} {Dropwise
  condensation theory and experiment: A review},}\ }\href {\doibase
  10.1243/09576500260049034} {\bibfield  {journal} {\bibinfo  {journal}
  {Proceedings of the Institution of Mechanical Engineers, Part A: Journal of
  Power and Energy}\ }\textbf {\bibinfo {volume} {216}},\ \bibinfo {pages}
  {115--128} (\bibinfo {year} {2002})}\BibitemShut {NoStop}%
\bibitem [{\citenamefont {{The OpenFOAM Foundation}}()}]{OpenFOAM}%
  \BibitemOpen
  \bibfield  {author} {\bibinfo {author} {\bibnamefont {{The OpenFOAM
  Foundation}}},\ }\href {www.openfoam.org} {\enquote {\bibinfo {title}
  {www.openfoam.org},}\ }\BibitemShut {NoStop}%
\bibitem [{\citenamefont {Manhart}(2004)}]{Manhart2004}%
  \BibitemOpen
  \bibfield  {author} {\bibinfo {author} {\bibfnamefont {M.}~\bibnamefont
  {Manhart}},\ }\bibfield  {title} {\enquote {\bibinfo {title} {A zonal grid
  algorithm for {DNS} of turbulent boundary layers},}\ }\href {\doibase
  10.1016/s0045-7930(03)00061-6} {\bibfield  {journal} {\bibinfo  {journal}
  {Computers {\&} Fluids}\ }\textbf {\bibinfo {volume} {33}},\ \bibinfo {pages}
  {435--461} (\bibinfo {year} {2004})}\BibitemShut {NoStop}%
\bibitem [{\citenamefont {Chorin}(1968)}]{Chorin1968}%
  \BibitemOpen
  \bibfield  {author} {\bibinfo {author} {\bibfnamefont {A.~J.}\ \bibnamefont
  {Chorin}},\ }\bibfield  {title} {\enquote {\bibinfo {title} {Numerical
  solution of the navier--stokes equations},}\ }\href {\doibase
  10.1090/S0025-5718-1968-0242392-2} {\bibfield  {journal} {\bibinfo  {journal}
  {Mathematics of Computation}\ }\textbf {\bibinfo {volume} {22}},\ \bibinfo
  {pages} {745--745} (\bibinfo {year} {1968})}\BibitemShut {NoStop}%
\bibitem [{\citenamefont {Kath}\ and\ \citenamefont {Wagner}(2016)}]{Kath2016}%
  \BibitemOpen
  \bibfield  {author} {\bibinfo {author} {\bibfnamefont {C.}~\bibnamefont
  {Kath}}\ and\ \bibinfo {author} {\bibfnamefont {C.}~\bibnamefont {Wagner}},\
  }\bibfield  {title} {\enquote {\bibinfo {title} {Highly resolved simulations
  of turbulent mixed convection in a vertical plane channel},}\ }in\ \href
  {\doibase 10.1007/978-3-319-27279-5_45} {\emph {\bibinfo {booktitle} {Notes
  on Numerical Fluid Mechanics and Multidisciplinary Design}}}\ (\bibinfo
  {publisher} {Springer International Publishing},\ \bibinfo {year} {2016})\
  pp.\ \bibinfo {pages} {515--524}\BibitemShut {NoStop}%
\bibitem [{\citenamefont {Lund}, \citenamefont {Wu},\ and\ \citenamefont
  {Squires}(1998)}]{Lund1998}%
  \BibitemOpen
  \bibfield  {author} {\bibinfo {author} {\bibfnamefont {T.~S.}\ \bibnamefont
  {Lund}}, \bibinfo {author} {\bibfnamefont {X.}~\bibnamefont {Wu}}, \ and\
  \bibinfo {author} {\bibfnamefont {K.~D.}\ \bibnamefont {Squires}},\
  }\bibfield  {title} {\enquote {\bibinfo {title} {Generation of turbulent
  inflow data for spatially-developing boundary layer simulations},}\ }\href
  {\doibase 10.1006/jcph.1998.5882} {\bibfield  {journal} {\bibinfo  {journal}
  {Journal of Computational Physics}\ }\textbf {\bibinfo {volume} {140}},\
  \bibinfo {pages} {233--258} (\bibinfo {year} {1998})}\BibitemShut {NoStop}%
\bibitem [{\citenamefont {Bellec}, \citenamefont {Toutant},\ and\ \citenamefont
  {Olalde}(2017)}]{Bellec2017}%
  \BibitemOpen
  \bibfield  {author} {\bibinfo {author} {\bibfnamefont {M.}~\bibnamefont
  {Bellec}}, \bibinfo {author} {\bibfnamefont {A.}~\bibnamefont {Toutant}}, \
  and\ \bibinfo {author} {\bibfnamefont {G.}~\bibnamefont {Olalde}},\
  }\bibfield  {title} {\enquote {\bibinfo {title} {Large eddy simulations of
  thermal boundary layer developments in a turbulent channel flow under
  asymmetrical heating},}\ }\href {\doibase 10.1016/j.compfluid.2016.07.001}
  {\bibfield  {journal} {\bibinfo  {journal} {Computers and Fluids}\ }\textbf
  {\bibinfo {volume} {151}},\ \bibinfo {pages} {159--176} (\bibinfo {year}
  {2017})}\BibitemShut {NoStop}%
\bibitem [{\citenamefont {Chan-Braun}, \citenamefont {Garc\'{i}a-Villalba},\
  and\ \citenamefont {Uhlmann}(2011)}]{Chan-Braun2011}%
  \BibitemOpen
  \bibfield  {author} {\bibinfo {author} {\bibfnamefont {C.}~\bibnamefont
  {Chan-Braun}}, \bibinfo {author} {\bibfnamefont {M.}~\bibnamefont
  {Garc\'{i}a-Villalba}}, \ and\ \bibinfo {author} {\bibfnamefont
  {M.}~\bibnamefont {Uhlmann}},\ }\bibfield  {title} {\enquote {\bibinfo
  {title} {Force and torque acting on particles in a transitionally rough
  open-channel flow},}\ }\href {\doibase 10.1017/jfm.2011.311} {\bibfield
  {journal} {\bibinfo  {journal} {Journal of Fluid Mechanics}\ }\textbf
  {\bibinfo {volume} {684}},\ \bibinfo {pages} {441--474} (\bibinfo {year}
  {2011})}\BibitemShut {NoStop}%
\bibitem [{\citenamefont {Köthe}, \citenamefont {Herzog},\ and\ \citenamefont
  {Wagner}(2014)}]{Koethe2014}%
  \BibitemOpen
  \bibfield  {author} {\bibinfo {author} {\bibfnamefont {T.}~\bibnamefont
  {Köthe}}, \bibinfo {author} {\bibfnamefont {S.}~\bibnamefont {Herzog}}, \
  and\ \bibinfo {author} {\bibfnamefont {C.}~\bibnamefont {Wagner}},\
  }\bibfield  {title} {\enquote {\bibinfo {title} {Shape optimization of
  aircraft cabin ventilation components using adjoint {CFD}},}\ }in\ \href
  {\doibase 10.1201/b17488-120} {\emph {\bibinfo {booktitle} {Engineering
  Optimization 2014}}}\ (\bibinfo  {publisher} {{CRC} Press},\ \bibinfo {year}
  {2014})\ pp.\ \bibinfo {pages} {675--680}\BibitemShut {NoStop}%
\bibitem [{\citenamefont {Kuerten}, \citenamefont {van~der Geld},\ and\
  \citenamefont {Geurts}(2011)}]{Kuerten2011}%
  \BibitemOpen
  \bibfield  {author} {\bibinfo {author} {\bibfnamefont {J.~G.~M.}\
  \bibnamefont {Kuerten}}, \bibinfo {author} {\bibfnamefont {C.~W.~M.}\
  \bibnamefont {van~der Geld}}, \ and\ \bibinfo {author} {\bibfnamefont
  {B.~J.}\ \bibnamefont {Geurts}},\ }\bibfield  {title} {\enquote {\bibinfo
  {title} {Turbulence modification and heat transfer enhancement by inertial
  particles in turbulent channel flow},}\ }\href {\doibase 10.1063/1.3663308}
  {\bibfield  {journal} {\bibinfo  {journal} {Physics of Fluids}\ }\textbf
  {\bibinfo {volume} {23}},\ \bibinfo {pages} {123301} (\bibinfo {year}
  {2011})}\BibitemShut {NoStop}%
\bibitem [{\citenamefont {Grass}(1971)}]{Grass1971}%
  \BibitemOpen
  \bibfield  {author} {\bibinfo {author} {\bibfnamefont {A.~J.}\ \bibnamefont
  {Grass}},\ }\bibfield  {title} {\enquote {\bibinfo {title} {Structural
  features of turbulent flow over smooth and rough boundaries},}\ }\href
  {\doibase 10.1017/s0022112071002556} {\bibfield  {journal} {\bibinfo
  {journal} {Journal of Fluid Mechanics}\ }\textbf {\bibinfo {volume} {50}},\
  \bibinfo {pages} {233--255} (\bibinfo {year} {1971})}\BibitemShut {NoStop}%
\bibitem [{\citenamefont {Madras}\ and\ \citenamefont
  {McCoy}(2001)}]{Madras2001}%
  \BibitemOpen
  \bibfield  {author} {\bibinfo {author} {\bibfnamefont {G.}~\bibnamefont
  {Madras}}\ and\ \bibinfo {author} {\bibfnamefont {B.~J.}\ \bibnamefont
  {McCoy}},\ }\bibfield  {title} {\enquote {\bibinfo {title} {Distribution
  kinetics theory of ostwald ripening},}\ }\href {\doibase 10.1063/1.1403687}
  {\bibfield  {journal} {\bibinfo  {journal} {The Journal of Chemical Physics}\
  }\textbf {\bibinfo {volume} {115}},\ \bibinfo {pages} {6699--6706} (\bibinfo
  {year} {2001})}\BibitemShut {NoStop}%
\bibitem [{\citenamefont {Kasagi}\ and\ \citenamefont
  {Nishimura}(1997)}]{Kasagi1997}%
  \BibitemOpen
  \bibfield  {author} {\bibinfo {author} {\bibfnamefont {N.}~\bibnamefont
  {Kasagi}}\ and\ \bibinfo {author} {\bibfnamefont {M.}~\bibnamefont
  {Nishimura}},\ }\bibfield  {title} {\enquote {\bibinfo {title} {Direct
  numerical simulation of combined forced and natural turbulent convection in a
  vertical plane channel},}\ }\href {\doibase 10.1016/S0142-727X(96)00148-8}
  {\bibfield  {journal} {\bibinfo  {journal} {International Journal of Heat and
  Fluid Flow}\ }\textbf {\bibinfo {volume} {18}},\ \bibinfo {pages} {88--99}
  (\bibinfo {year} {1997})}\BibitemShut {NoStop}%
\bibitem [{\citenamefont {Wetzel}\ and\ \citenamefont
  {Wagner}(2019)}]{Wetzel2019}%
  \BibitemOpen
  \bibfield  {author} {\bibinfo {author} {\bibfnamefont {T.}~\bibnamefont
  {Wetzel}}\ and\ \bibinfo {author} {\bibfnamefont {C.}~\bibnamefont
  {Wagner}},\ }\bibfield  {title} {\enquote {\bibinfo {title} {Buoyancy-induced
  effects on large-scale motions in differentially heated vertical channel
  flows studied in direct numerical simulations},}\ }\href {\doibase
  10.1016/j.ijheatfluidflow.2018.09.005} {\bibfield  {journal} {\bibinfo
  {journal} {International Journal of Heat and Fluid Flow}\ }\textbf {\bibinfo
  {volume} {75}},\ \bibinfo {pages} {14--26} (\bibinfo {year}
  {2019})}\BibitemShut {NoStop}%
\end{thebibliography}%

\end{document}